\documentclass[letterpaper,12pt]{article}

\usepackage{mathtools,amssymb,amsmath,amsthm}
\usepackage{graphicx}
\usepackage{enumerate}
\usepackage{mathrsfs}
\usepackage[mathscr]{eucal}
\usepackage{booktabs}
\usepackage{subcaption}

\usepackage{lineno}
\graphicspath{{figs/}}

\usepackage[letterpaper, margin=1in]{geometry}
\usepackage[colorlinks,linkcolor=blue,citecolor=blue]{hyperref}
\usepackage{cleveref}

\newtheorem{theorem}{Theorem}

\newtheorem{observation}{Observation}

\DeclareMathOperator{\polylog}{polylog}

\newcommand{\tmin}{t_{\min}}

\Crefname{lemma}{Lemma}{Lemmas}
\Crefname{table}{Table}{Tables}
\Crefname{observation}{Observation}{Observations}
\Crefname{fact}{Fact}{Facts}
\Crefname{corollary}{Corollary}{Corollaries}

\title{Trajectory Visibility at First Sight\thanks{A preliminary version of this work appeared in the Proceedings of the 20th International Conference and Workshops on Algorithms and Computation (WALCOM 2026), LNCS 16444, pp.~499--511, Springer, 2026.}}

\author{
    Mohammad Ali Abam\thanks{Department of Computer Engineering, Sharif University of Technology, Tehran, Iran. Email: \texttt{abam@sharif.edu}. Authors' names are sorted alphabetically. Corresponding author.}
    \and
    Mohammad Ghodsi\thanks{Department of Computer Engineering, Sharif University of Technology, Tehran, Iran; and School of Computer Science, Institute for Research in Fundamental Sciences (IPM), Tehran, Iran. Email: \texttt{ghodsi@sharif.edu}.}
    \and
    Seyed Mohammad Hussein Kazemi\thanks{Department of Computer Engineering, Sharif University of Technology, Tehran, Iran; and LaBRI, University of Bordeaux, Bordeaux, France. Email: \texttt{skazemi@u-bordeaux.fr}.}
}

\date{}

\begin{document}

\maketitle
\pagestyle{plain}


\begin{abstract}
Let $P$ be a simple polygon with $n$ vertices, and let two moving entities $q(t)$ and $r(t)$ travel at constant (possibly distinct) speeds $v_q$ and $v_r$ along line-segment trajectories $\tau_q$ and $\tau_r$ inside $P$. We study the first-visibility time
$
t^* = \min\bigl\{\,t \ge 0 \mid \overline{q(t)\,r(t)} \subseteq P\,\bigr\},
$
the earliest moment at which the segment joining $q(t)$ and $r(t)$ lies entirely within $P$. Prior work by Eades et al.~\cite{swat2020,eeades-phd} focused on the decision version of this question in a simple polygon. They gave a one-shot decision algorithm running in $\mathcal{O}(n)$ time. For a stationary entity and a moving one, they gave a data structure with $\mathcal{O}(n \log n)$ preprocessing time and $\mathcal{O}(n)$ space that answers decision queries in $\mathcal{O}(\log n)$ time. For two moving entities, after $\mathcal{O}(n \log^5 n)$ preprocessing time, they built a data structure of size $\mathcal{O}(n \log^5 n)$ with $\mathcal{O}(n^{3/4}\log^3 n)$ query time. Variants for polygonal domains with holes or when entities cross the boundary of $P$ lie beyond our scope.

In this work, we go beyond decision to compute $t^*$ exactly under three models for a simple polygon $P$. When both trajectories are fixed in advance, we preprocess $P$ in $\mathcal{O}(n)$ time and space and answer each query $(v_q, v_r)$ in $\mathcal{O}(\log n)$ time. If one trajectory $\tau_r$ is fixed while $\tau_q$ is given as a query, we build a structure in $\mathcal{O}(n \log n)$ time and space that computes $t^*$ in $\mathcal{O}(\log^2 n)$ time per query. When neither trajectory is known in advance, we develop a data structure with $\mathcal{O}(n^{1+\varepsilon})$ expected preprocessing time and $\mathcal{O}(n)$ space that achieves $\mathcal{O}(n^{1/2}\,\polylog(n))$ expected query time for any fixed $\varepsilon > 0$.
\end{abstract}



\section{Introduction}

The analysis of moving entities constitutes a broad research area, ranging from robotics and geographic information science (GIScience) to meteorology and ecology~\cite{stohl1998computation, calenge2009concept, gurarie2009novel, li2010deriving, dodge2009revealing}. A body of work focuses on extracting features from historical trajectory data, such as identifying when entities have passed closely together or detecting groups that move in formation~\cite{11475/15060}. However, a class of problems emerges when the goal is not to analyse the past, but to predict future events for entities on known or planned paths. This is especially true in environments with obstacles or defined boundaries, which mirrors the constrained polygonal setting of our work.

Practical applications of such predictive line-of-sight analysis are numerous. In robotics and automated systems, for example, determining when two microrobots can establish direct communication~\cite{pane2021real} or when an autonomous vehicle will gain a clear view of a target is a fundamental task. Similar challenges arise in biology when tracking potential interactions within animal colonies in complex habitats~\cite{bozek2021markerless} and in optimising sensor-based systems, such as dual-axis solar trackers that must maintain an unobstructed path to the sun~\cite{jamroen2021novel}. As the complexity and autonomy of these systems grow, the demand for a robust theory of trajectory visibility and the development of efficient, provable algorithms naturally escalates.

\subsection{Preliminaries}

Throughout this paper, we work in the standard real RAM model (as in~\cite{agarwal2013range}), augmented with standard unit-cost word-RAM operations on $\mathcal{O}(\log n)$-bit words, such as bitwise logical operations and table lookups (as in~\cite{chazelle1988functional}).

Let $P$ be a simple polygon with $n$ vertices, and let $\partial P$ denote the boundary of $P$, consisting of its vertices and edges. A point $p \in P$ is \emph{weakly visible} from a segment $s \subset P$ if there exists a point $q \in s$ such that $\overline{pq} \subseteq P$. Similarly, two segments $s_1, s_2 \subset P$ are weakly visible to each other if for every point in one segment there exists a visible point in the other~\cite{ghosh2007}. For a query point $p \in P$, the \emph{visibility polygon} of $p$ is $V(p) = \{q \in P \mid \overline{pq} \subseteq P\}$, computable in $\mathcal{O}(n)$ time~\cite{lee1983visibility,joe1987corrections}. Given a segment $s \subset P$, the \emph{weak visibility polygon} of $s$ is $W(s) = \{q \in P \mid \exists\, r \in s \text{ such that } \overline{qr} \subseteq P\}$, computable in $\mathcal{O}(n)$ time~\cite{ghosh2007}; its complement $P \setminus W(s)$ consists of pairwise disjoint pockets, each separated from $W(s)$ by a single straight \emph{window} of $\partial W(s)$~\cite{ghosh2007}. A \emph{ray-shooting} query asks, for a query ray $\overrightarrow{r}$ with origin $p \in P$, for the first intersection of $\overrightarrow{r}$ with $\partial P$; after $\mathcal{O}(n)$ preprocessing time and space, such queries are answered in $\mathcal{O}(\log n)$ time~\cite{chazelle1994ray}.

A splinegon $\mathcal{S}$ is formed from a simple polygon by replacing one or more of its edges with curved edges so that the region enclosed by each curved edge and its chord is convex~\cite{dobkin1990computational}. Given a simple splinegon $\mathcal{S}$ with $n$ edges, there exists an $\mathcal{O}(n)$-space data structure built in $\mathcal{O}(n)$ time that reports the first intersection of any query ray $\overrightarrow{r}$ emanating from a point $p \in \mathcal{S}$ with $\partial \mathcal{S}$ in $\mathcal{O}(\log n)$ time~\cite{melissaratos1992shortest,GHODSI2014435}.

A simple polygon $P$ with $n$ vertices can be triangulated in $\mathcal{O}(n)$ time~\cite{chazelle1991triangulating}. Recursively splitting the dual tree of the triangulation via balanced diagonal cuts, while splitting any subpolygon bounded by three diagonals at the unique triangle separating them~\cite{guibas1989optimal,chazelle1994ray}, yields a balanced hierarchy $\mathcal{D}$ of maximum degree $3$ and height $h \in \mathcal{O}(\log n)$. Every node $u \in \mathcal{D}$ corresponds to a subpolygon $P_u \subseteq P$ bounded by at most three diagonals, and the path between any two triangles in the dual tree can be decomposed into $\mathcal{O}(\log n)$ interior-disjoint subpolygons of $\mathcal{D}$.

Consider a point $p \in P$. For every vertex $v$ of $P$, let $\pi(p, v)$ denote the shortest Euclidean path inside $P$ connecting $p$ and $v$. The shortest-path tree rooted at $p$, denoted as $T(p) = \bigcup_{v \in P} \pi(p, v)$, has all intermediate vertices on each $\pi(p, v)$ as reflex vertices of $P$~\cite{aronov2002visibility}. Extending each edge $(u, v)$ of $T(p)$ past $v$ until it hits $\partial P$ partitions $P$ into $\mathcal{O}(n)$ regions, called the \emph{shortest-path map} of $p$, such that all points $q$ in a given region share the same parent vertex (either $p$ or a reflex vertex of $P$) along $\pi(p, q)$. Refining these regions into triangles yields an $\mathcal{O}(n)$-size triangular subdivision in $\mathcal{O}(n)$ time~\cite{chazelle1991triangulating,guibas1987linear}, which can be preprocessed in $\mathcal{O}(n)$ time and space for $\mathcal{O}(\log n)$-time point location~\cite{kirkpatrick1983optimal} to locate the parent of any query point $q \in P$ in $T(p)$.

For a line segment $s \subset P$, we denote its endpoints by $p^+_s$ and $p^-_s$; when considering two segments $s_1, s_2 \subset P$ whose endpoints lie in convex position, $p^+_{s_1}, p^+_{s_2}$ and $p^-_{s_1}, p^-_{s_2}$ denote the corresponding pairs of endpoints on the boundary of their convex hull. For any point $a \in P$ and segment $s \subset P$, the union of shortest paths $\pi(a, b)$ over all $b \in s$ forms a \emph{funnel} $\mathcal{F}(a, s)$~\cite{guibas1989optimal}: the paths $\pi(a, p^+_s)$ and $\pi(a, p^-_s)$ share a common prefix (the \emph{stem}) from $a$ to a vertex $c$ (the \emph{apex}), and then diverge into two inward-convex chains $\pi(c, p^+_s)$ and $\pi(c, p^-_s)$ to the endpoints of $s$. More generally, for two segments $s_1, s_2 \subset P$, their \emph{hourglass} $H(s_1, s_2)$ (the union of all shortest paths between points of $s_1$ and $s_2$) is either \emph{open} (bounded by two disjoint inward-convex chains $\gamma^+ = \pi(p^+_{s_1}, p^+_{s_2})$ and $\gamma^- = \pi(p^-_{s_1}, p^-_{s_2})$) or \emph{closed} (two funnels joined by a central shortest-path stem between their apexes)~\cite{guibas1989optimal}. Their \emph{visibility glass} $L(s_1, s_2)$ is the (potentially empty) union of all line segments between $s_1$ and $s_2$ that lie within $P$. When $L(s_1, s_2)$ is non-empty, there exist subsegments $s'_1 \subseteq s_1$ and $s'_2 \subseteq s_2$ bounded by two inner bi-tangents along the cross-paths $\pi(p^+_{s_1}, p^-_{s_2})$ and $\pi(p^-_{s_1}, p^+_{s_2})$ such that $L(s_1, s_2) = H(s'_1, s'_2)$ (computable in $\mathcal{O}(n)$ time~\cite{swat2020}), and each cross-path switches turn direction at most once (Observation~1 of~\cite{swat2020}).

Over the hierarchy $\mathcal{D}$, Guibas and Hershberger~\cite{guibas1989optimal} build an $\mathcal{O}(n)$-space two-point shortest-path structure in $\mathcal{O}(n)$ time (after triangulation) answering shortest-path queries in $\mathcal{O}(\log n)$ time. Fundamental hourglasses of subpolygons are built persistently bottom-up via $\mathcal{O}(\log n)$-time chain concatenation, representing each convex chain as a binary tree of height $\mathcal{O}(\log n)$ whose internal nodes $v$ store a common-tangent edge $e_v$ along with left and right edge-retrieval lists (Lemmas~4--6 of~\cite{guibas1989optimal}). Because concatenating two chains trims away the subchains tucked behind their common tangent without modifying the original subtrees, an edge $e_v$ deeper in the tree can lie outside the surviving subpath along $\partial P$ (or collapse to a zero-length tangent at a shared vertex); for any surviving subpath spanning a boundary index interval $[a, b]$ along $\partial P$, one can test in $\mathcal{O}(1)$ time whether $e_v$ falls inside $[a, b]$ (Section~4.2 of~\cite{guibas1989optimal}) and replace a zero-length tangent edge by its adjacent edge from the retrieval list. Using this structure, Eades et al.~\cite{swat2020} (Lemma~8) decide whether $L(s_1, s_2)$ is empty or compute $H(s'_1, s'_2)$ in $\mathcal{O}(\log n)$ time.

In their work on semi-algebraic range searching, Agarwal et al.~\cite{agarwal2013range} showed that for fixed constants $d, \Delta, s$ and any $\varepsilon > 0$, an arbitrary $n$-point set in $\mathbb{R}^d$ admits a partition tree with $\mathcal{O}(n^{1+\varepsilon})$ \emph{expected} preprocessing time and $\mathcal{O}(n)$ space that answers any constant-complexity semi-algebraic range query in $\mathcal{O}(n^{1-1/d}\log^{B_0} n)$ \emph{expected} time, where $B_0 = B_0(d, \Delta, s, \varepsilon)$.\footnote{Under $D_0$-general position (Theorem~1.2 of~\cite{agarwal2013range}), the expected preprocessing time improves to $\mathcal{O}(n \log n)$ with expected query time $\mathcal{O}(n^{1-1/d+\varepsilon})$.} In their $d = 2$ construction (Sections~6 and~7 of~\cite{agarwal2013range}), every input point lies in a unique leaf, either of a 2D partition subtree or of a 1D binary search tree on an exceptional monotone algebraic arc, so an in-order leaf traversal yields a single permutation in which each of the $\mathcal{O}(n^{1/2}\log^{B_0} n)$ canonical subsets or 1D intervals returned by a query is a contiguous subarray.

For two-dimensional orthogonal range structures, Chazelle~\cite{chazelle1988functional} showed how to compress $n$ points on $\{1,\dots,n\}^2$ across $\mathcal{O}(\log n)$ binary range-tree levels into $\mathcal{O}(n)$ words of space in $\mathcal{O}(n \log n)$ time via bit-packed split vectors supporting $\mathcal{O}(1)$-time prefix-count queries.

For a 3D convex polyhedron with $m$ vertices, the Dobkin--Kirkpatrick hierarchy~\cite{dobkin1990determining} uses $\mathcal{O}(m)$ space and preprocessing time and answers halfspace-emptiness queries in $\mathcal{O}(\log m)$ time; moreover, by point-plane duality, two 3D convex hulls can be merged in $\mathcal{O}(m)$ time~\cite{chazelle1992optimal}. Given an ordered sequence of items with positive weights $w_1, \dots, w_k$ and total weight $W = \sum_{i=1}^k w_i$, a \emph{biased binary tree} is a binary search tree storing the items in order such that each item $i$ lies at depth $\mathcal{O}(\log(W / w_i))$~\cite{sleator1983dynamic}.

Consider a rooted tree $T$ with $n$ vertices. Sleator and Tarjan~\cite{sleator1983dynamic} partition $T$ into vertex-disjoint \emph{solid paths} by calling an edge $(v, \operatorname{parent}(v))$ solid if $2 \cdot \operatorname{size}(v) > \operatorname{size}(\operatorname{parent}(v))$, and off-path otherwise. Each vertex $v \in T$ is given a weight equal to $1$ plus the total size of the subtrees rooted at its off-path children, and each solid path is stored as a biased binary tree with these weights. Then every vertex has depth $\mathcal{O}(\log n)$ in its biased binary tree. Moreover, any path in $T$ from an ancestor to a descendant can be decomposed into $\mathcal{O}(\log n)$ canonical nodes of these trees and $\mathcal{O}(\log n)$ off-path edges in path order~\cite{sleator1983dynamic}.

Let a trajectory be a sequence of time-stamped locations in $\mathbb{R}^2$ modeling the movement of an entity in a polygon. The trajectory visibility problem was studied by Eades et al.~\cite{swat2020}, who asked whether there is ever a time at which two moving entities $q$ and $r$ can see each other. Throughout this paper, $q$ and $r$ move along line-segment trajectories $\tau_q, \tau_r \subset P$ at constant (possibly distinct) speeds $v_q$ and $v_r$ and cannot see through edges of $P$. Let $t_{qf}$ and $t_{rf}$ denote the times required for $q$ and $r$ to traverse $\tau_q$ and $\tau_r$, respectively. If one entity reaches the end of its segment at time $\tmin = \min(t_{qf}, t_{rf})$ before the other finishes at $t_f = \max(t_{qf}, t_{rf})$, the motion splits into two constant-speed phases $[0, \tmin]$ (both moving) and $[\tmin, t_f]$ (one stationary with speed zero and one moving). Handling the two phases independently (returning the answer from the first phase if visible, and otherwise checking the second), we assume below that both entities move at constant speeds throughout the considered time interval, with positions:
$$
q(t) = (x_q(t), y_q(t)) = (x_{q,0} + v_{qx} t, y_{q,0} + v_{qy} t)
$$
\noindent and: 
$$
r(t) = (x_r(t), y_r(t)) = (x_{r,0} + v_{rx} t, y_{r,0} + v_{ry} t)
$$
where $(x_{q,0}, y_{q,0})$ and $(x_{r,0}, y_{r,0})$ are their positions at the start of the phase (shifting time to $t = 0$ for the second phase) and $(v_{qx}, v_{qy}), (v_{rx}, v_{ry})$ are their speed vectors.

\begin{table}[!tp]
  \centering
  \caption{Preprocessing, storage, and query-time complexities for trajectory visibility in a simple polygon with line-segment trajectories. The first two rows list the best-known data-structure bounds from Eades et al.~\cite{swat2020} for deciding whether mutual visibility ever occurs (\textbf{Decision}). The last three rows summarise our results (Theorems~\ref{theorem::FixedTrajectories}, \ref{theorem::oneQueryTraj}, and~\ref{theorem::queryTrajs}), which compute the earliest visibility time $t^*$ (\textbf{Finding $t^*$}).}
  \label{tab:comparison}
  \begin{tabular}{lcccc}
    \toprule
    Reference & Scenario & Preprocessing & Space & Query time \\
    \midrule
    \cite{swat2020} (Sec.\,5)
      & Stationary \& moving
      & $\mathcal{O}(n \log n)$ & $\mathcal{O}(n)$
      & $\mathcal{O}(\log n)$ \\
      & (\textbf{Decision}) & & & \\

    \cite{swat2020} (Sec.\,4)
      & Query trajectories
      & $\mathcal{O}(n\log^{5}n)$
      & $\mathcal{O}(n\log^{5}n)$
      & $\mathcal{O}\bigl(n^{3/4}\log^{3}n\bigr)$  \\
      & (\textbf{Decision}) & & & \\

    Theorem~\ref{theorem::FixedTrajectories}
      & Fixed trajectories
      & $\mathcal{O}(n)$
      & $\mathcal{O}(n)$
      & $\mathcal{O}(\log n)$ \\
      & (\textbf{Finding $t^*$}) & & & \\

    Theorem~\ref{theorem::oneQueryTraj}
      & One trajectory as query
      & $\mathcal{O}(n \log n)$
      & $\mathcal{O}(n \log n)$
      & $\mathcal{O}(\log^2 n)$ \\
      & (\textbf{Finding $t^*$}) & & & \\

    Theorem~\ref{theorem::queryTrajs}
      & Query trajectories
      & $\mathcal{O}\bigl(n^{1+\varepsilon}\bigr)$
      & $\mathcal{O}(n)$
      & $\mathcal{O}\bigl(n^{1/2}\log^{B}n\bigr)$  \\
      & (\textbf{Finding $t^*$}) & (expected) & & (expected) \\

    \bottomrule
  \end{tabular}
\end{table}

\subsection{Our Contributions}

From now on, we consider a simple polygon $P$ and line-segment trajectories $\tau_q, \tau_r \subset P$ that never cross $\partial P$. We address directions left open in~\cite{swat2020,eeades-phd} by computing the earliest visibility time $t^*$ under three models (see Table~\ref{tab:comparison}):

\begin{enumerate}
    \item Suppose two fixed line-segment trajectories $\tau_q, \tau_r \subset P$ are given. We preprocess $P$ in $\mathcal{O}(n)$ time and space so that for any query speeds $v_q$ and $v_r$, the earliest visibility time $t^*$ is computed in $\mathcal{O}(\log n)$ time (Theorem~\ref{theorem::FixedTrajectories}).

    \item Suppose $\tau_r \subset P$ is fixed while $\tau_q \subset P$ is given as a query along with $v_q$ and $v_r$. We build a data structure in $\mathcal{O}(n \log n)$ time and space that computes $t^*$ in $\mathcal{O}(\log^2 n)$ time (Theorem~\ref{theorem::oneQueryTraj}).

    \item Finally, when neither $\tau_q$ nor $\tau_r$, nor $v_q$ and $v_r$, are known in advance, we construct a data structure in $\mathcal{O}(n^{1+\varepsilon})$ \emph{expected} preprocessing time and $\mathcal{O}(n)$ space that computes $t^*$ in $\mathcal{O}(n^{1/2}\log^{B} n)$ \emph{expected} time, where $B$ is a constant depending on $\varepsilon$ (Theorem~\ref{theorem::queryTrajs}).
\end{enumerate}

We note an earlier preliminary preprint by a subset of the authors~\cite{kazemi2022trajectory}; our current approach does not overlap with it and resolves serious issues in its analysis.


\section{Fixed Trajectories}
\label{sec::fixedTrajs}

\begin{figure}[t]
  \centering
  \begin{subfigure}[t]{0.48\textwidth}
      \centering
      \includegraphics[height=1.8in]{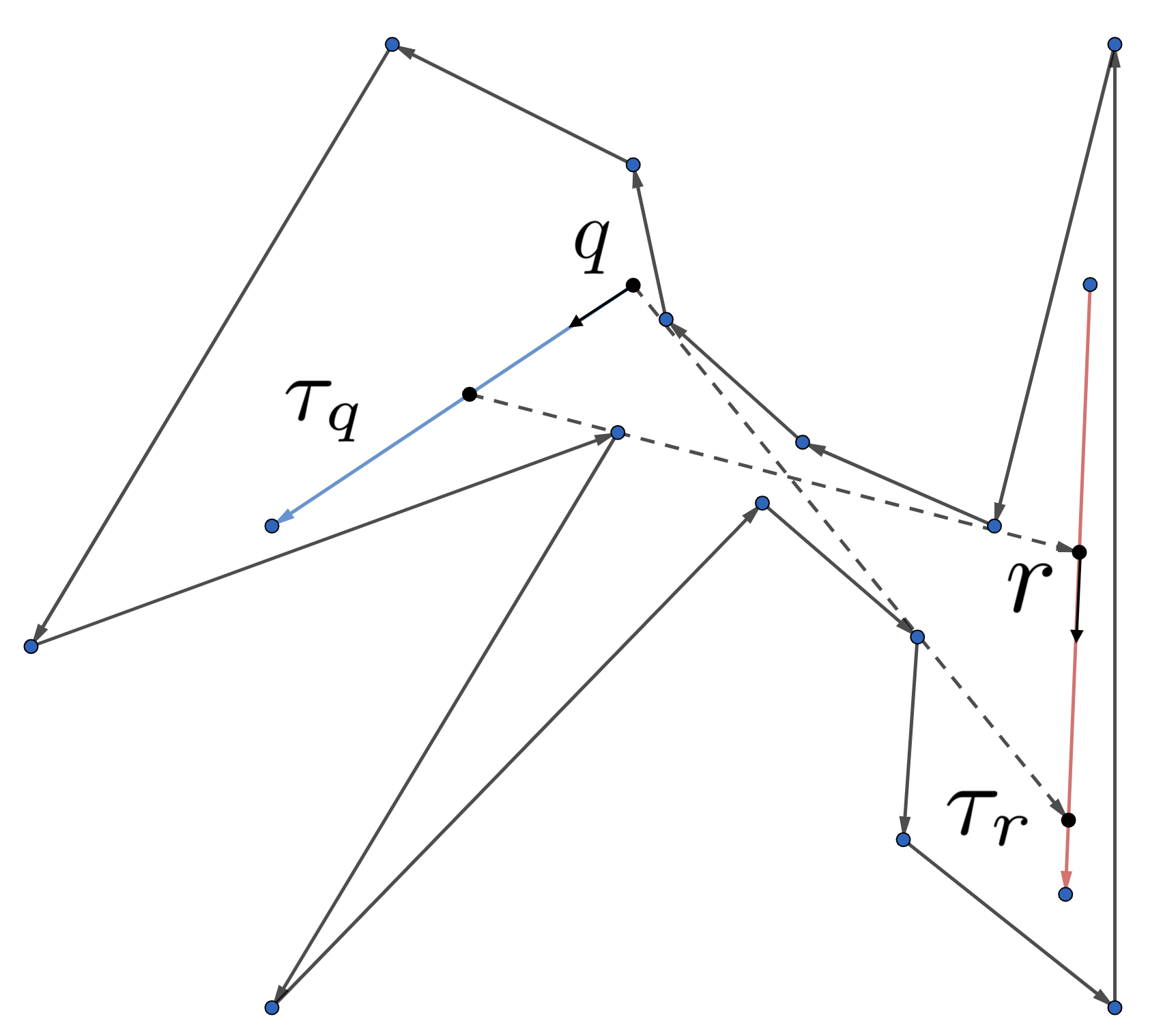}
  \end{subfigure}%
  ~
  \begin{subfigure}[t]{0.48\textwidth}
      \centering
      \includegraphics[height=1.8in]{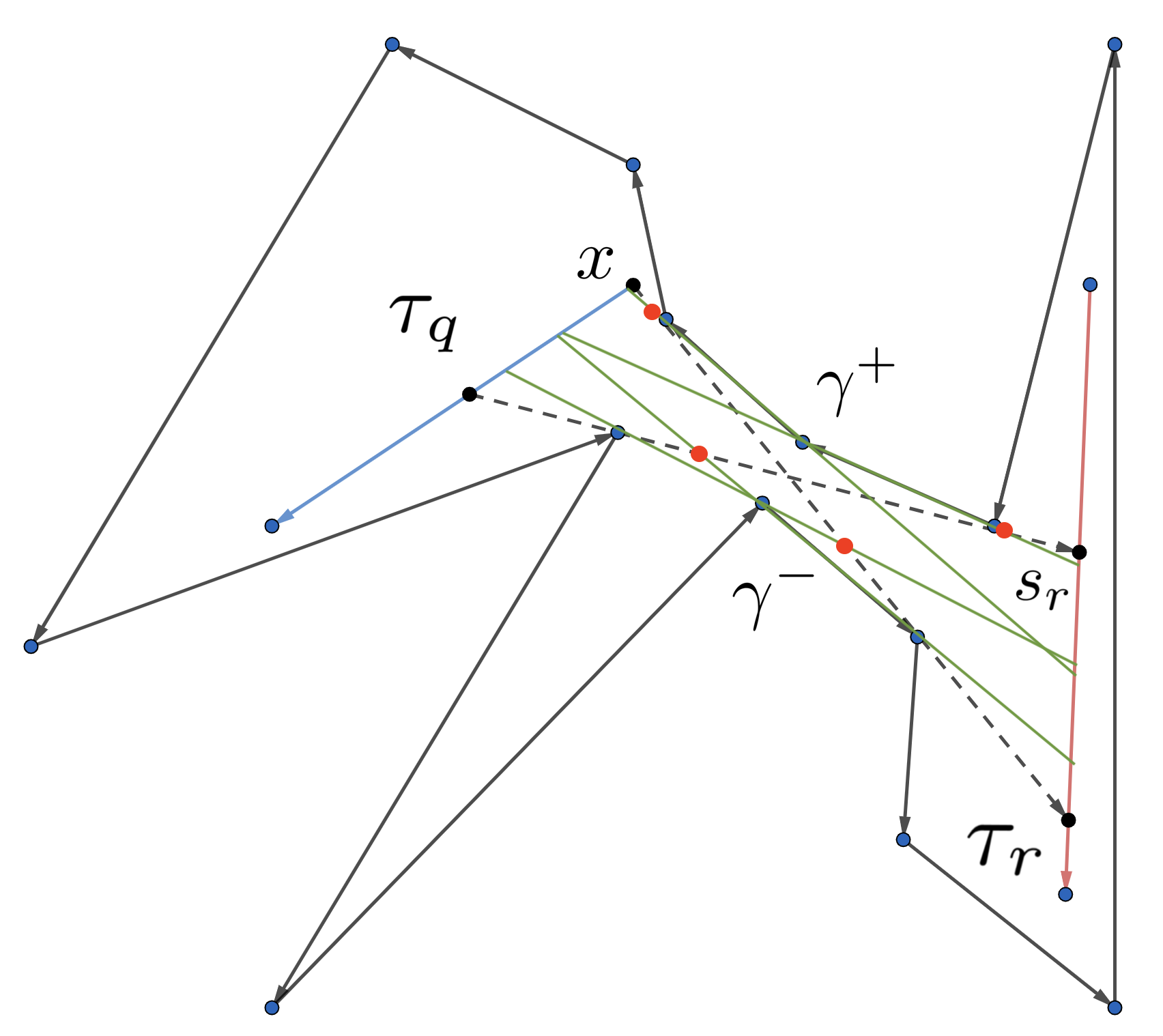}
  \end{subfigure}
  \caption{(Left) Given two moving entities $q$ and $r$ with their trajectories $\tau_r$ and $\tau_q$ in red and blue within a simple polygon, the dotted lines indicate the bi-tangents of their visibility glass. (Right) Extensions of segments between consecutive reflex vertices (in green) (known as critical constraints~\cite{aronov2002visibility}) intersect their bi-tangents (red dots). For every $x \in I_q$, there is a corresponding subsegment $s_r \subset \tau_r$ such that every point in $s_r$ is visible to $x$ within $P$.}
  \label{LAndExL}
\end{figure}

Consider the visibility glass $L(\tau_q, \tau_r)$, which can be constructed in $\mathcal{O}(n)$ time~\cite{swat2020}; if it is empty, no visibility occurs. For every pair of consecutive reflex vertices $v_i$ and $v_{i+1}$ on its upper chain $\gamma^+$ (and similarly on $\gamma^-$), let $l(v_i, v_{i+1})$ be the line through them. Let the ray emitting from $v_i$ to $v_{i+1}$ on $l(v_i, v_{i+1})$ refer to the ray with origin at $v_i$ directed along $l(v_i, v_{i+1})$ toward $v_{i+1}$ (similarly for the ray from $v_{i+1}$ to $v_i$). Let each of these rays be followed until its first intersection with $\partial P$ (excluding $v_i$ and $v_{i+1}$). The closed segment of $l(v_i, v_{i+1})$ between those two intersection points on $\partial P$ is denoted as a \emph{critical constraint} by Aronov et al.~\cite{aronov2002visibility} and others before them. Because each supporting line $l(v_i, v_{i+1})$ is unobstructed inside the visibility glass $L(\tau_q, \tau_r)$, its intersections with $\tau_q$ and $\tau_r$ are computed directly in $\mathcal{O}(1)$ time, taking $\mathcal{O}(n)$ time overall. Define:
\[
I_q = \bigcup_i \bigl(l(v_i, v_{i+1}) \cap \tau_q\bigr)
\quad\text{and}\quad
I_r = \bigcup_i \bigl(l(v_i, v_{i+1}) \cap \tau_r\bigr)
\]
including the endpoints of the two bi-tangents on $\tau_q$ and $\tau_r$. These intersection points partition $\tau_q$ and $\tau_r$ into subsegments. For each intersection point $x \in I_q$, there is a corresponding subsegment $s_r \subset \tau_r$ such that every point in $s_r$ is visible to $x$ within $P$ (see Figure~\ref{LAndExL}).

Next, we define a mapping to map positions along $\tau_q$ and $\tau_r$ into a two-dimensional diagram $D \subset \mathbb{R}^2$. Let $\varphi_q : \tau_q \to [0, \operatorname{len}(\tau_q)]$ and $\varphi_r : \tau_r \to [0, \operatorname{len}(\tau_r)]$ be the arc-length parametrizations of $\tau_q$ and $\tau_r$, respectively. Note that we map the starting point of the trajectories to $0$. In $D$, the $x$-coordinate is determined by $\varphi_q$ and the $y$-coordinate by $\varphi_r$. We define the diagram as:
\[
D = [0, \operatorname{len}(\tau_q)] \times [0, \operatorname{len}(\tau_r)] \subset \mathbb{R}^2
\]
A point $(x, y) \in D$ represents the pair $\varphi_q^{-1}(x)$ on $\tau_q$ and $\varphi_r^{-1}(y)$ on $\tau_r$. We call $(x, y)$ \emph{visible} if these two points are mutually visible in $P$. The set of visible points in $D$ is the mutual-visibility region $V \subset D$, and its boundary consists of parameter pairs $(x, y)$ at which visibility holds.

\begin{figure}[t]
\centering
\begin{subfigure}[t]{0.48\textwidth}
  \centering
  \includegraphics[height=2.2in]{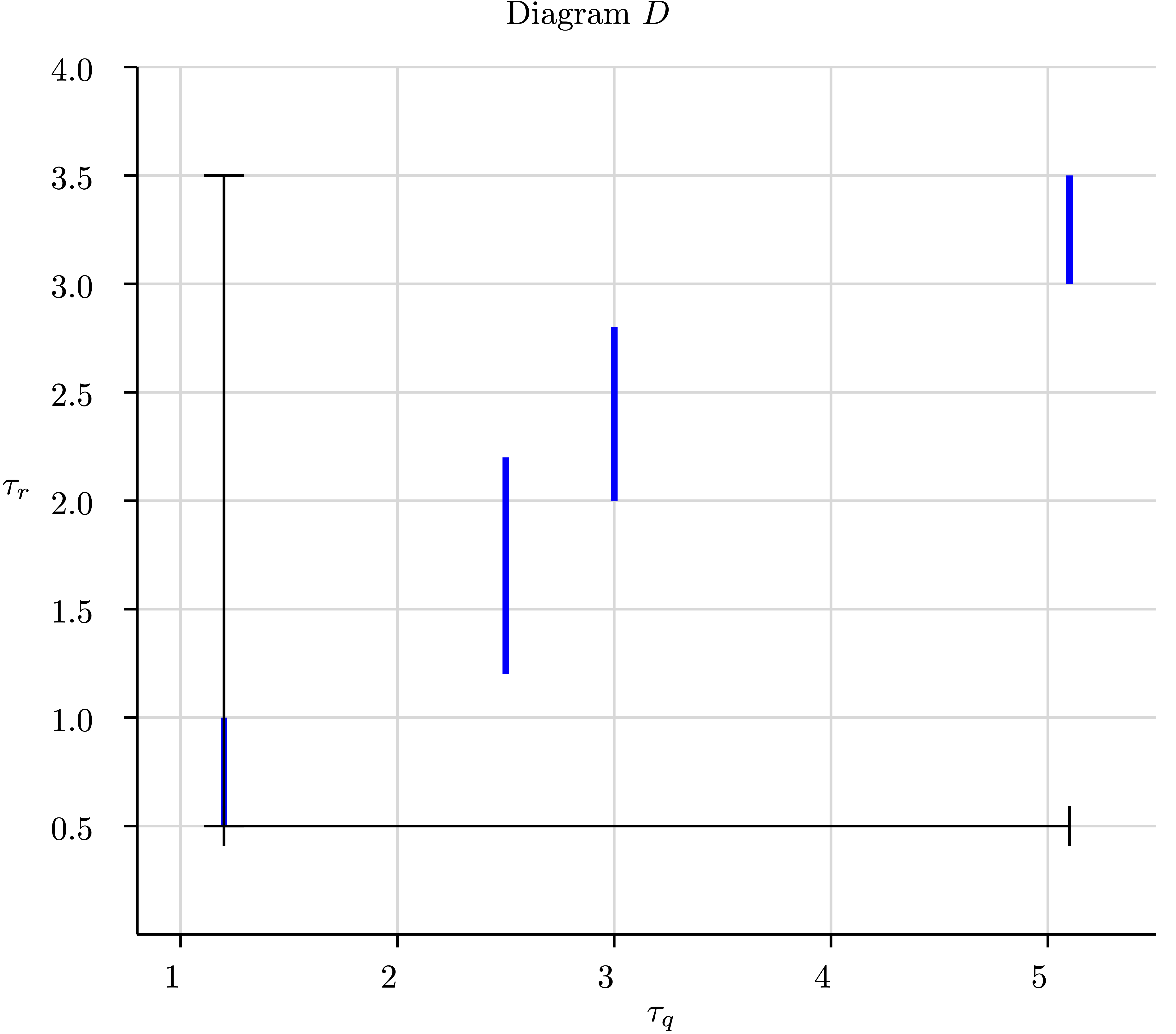}
\end{subfigure}%
~
\begin{subfigure}[t]{0.48\textwidth}
  \centering
  \includegraphics[height=2.2in]{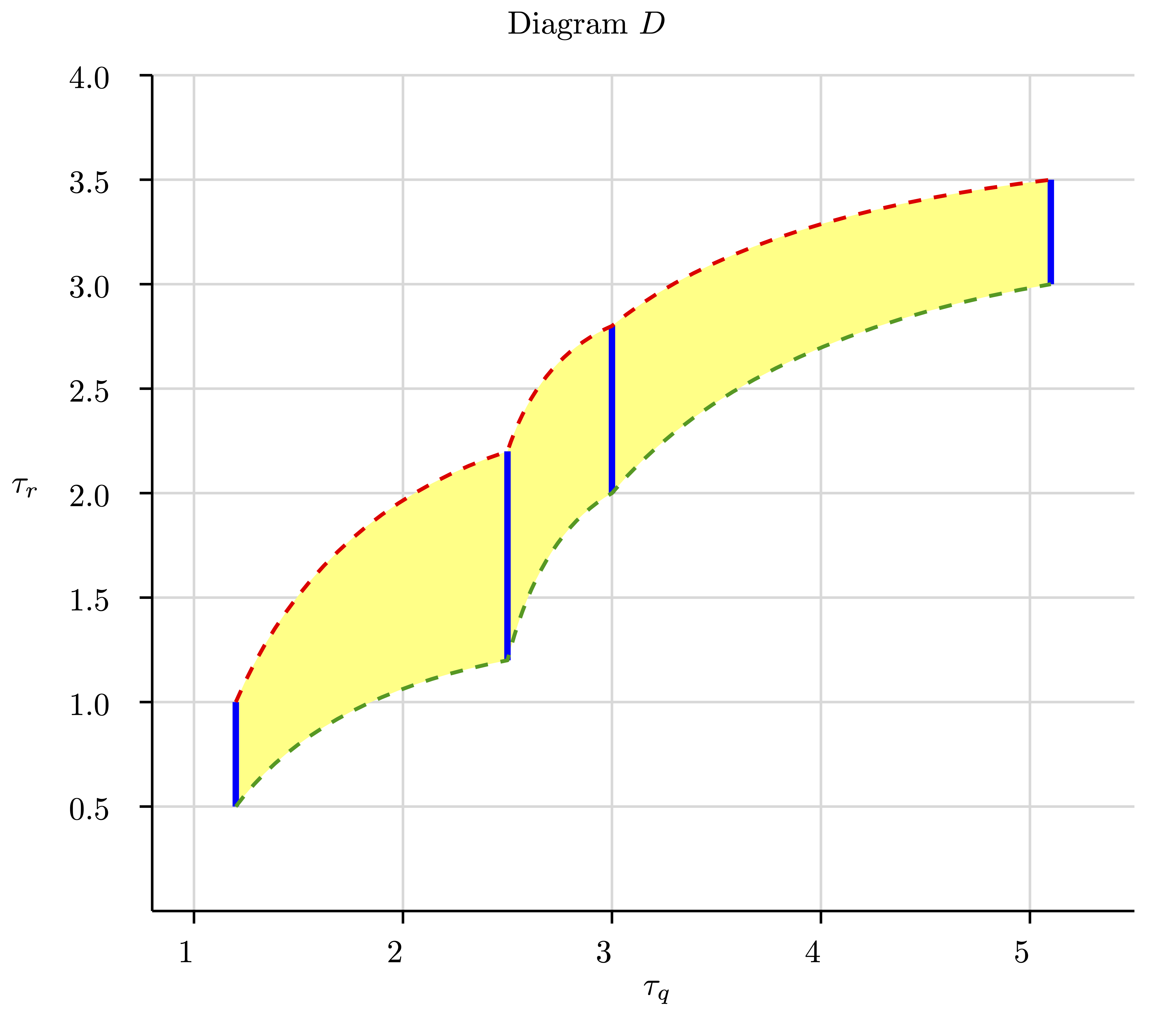}
\end{subfigure}
\caption{(Left) Vertical segments in the diagram $D$ induced by the intersection points $I_q$ on $\tau_q$: each $x=\varphi_q(x_i)$ spans the interval of $\varphi_r$-values for which points on $\tau_r$ are visible to $x_i$. (Right) The curved boundary (splinegon) obtained by the nonlinear mapping discussed in Observation~\ref{obs::mapOnOtherTraj} (Section~\ref{sec::fixedTrajs}).}
\label{DiagramD}
\end{figure}

Each $x \in I_q$ corresponds to a vertical segment in $D$ that represents a segment along $\tau_r$ on which every point is visible to $x$ (see Figure~\ref{DiagramD}).

\begin{figure}[t]
    \centering
    \includegraphics[width=0.45\linewidth]{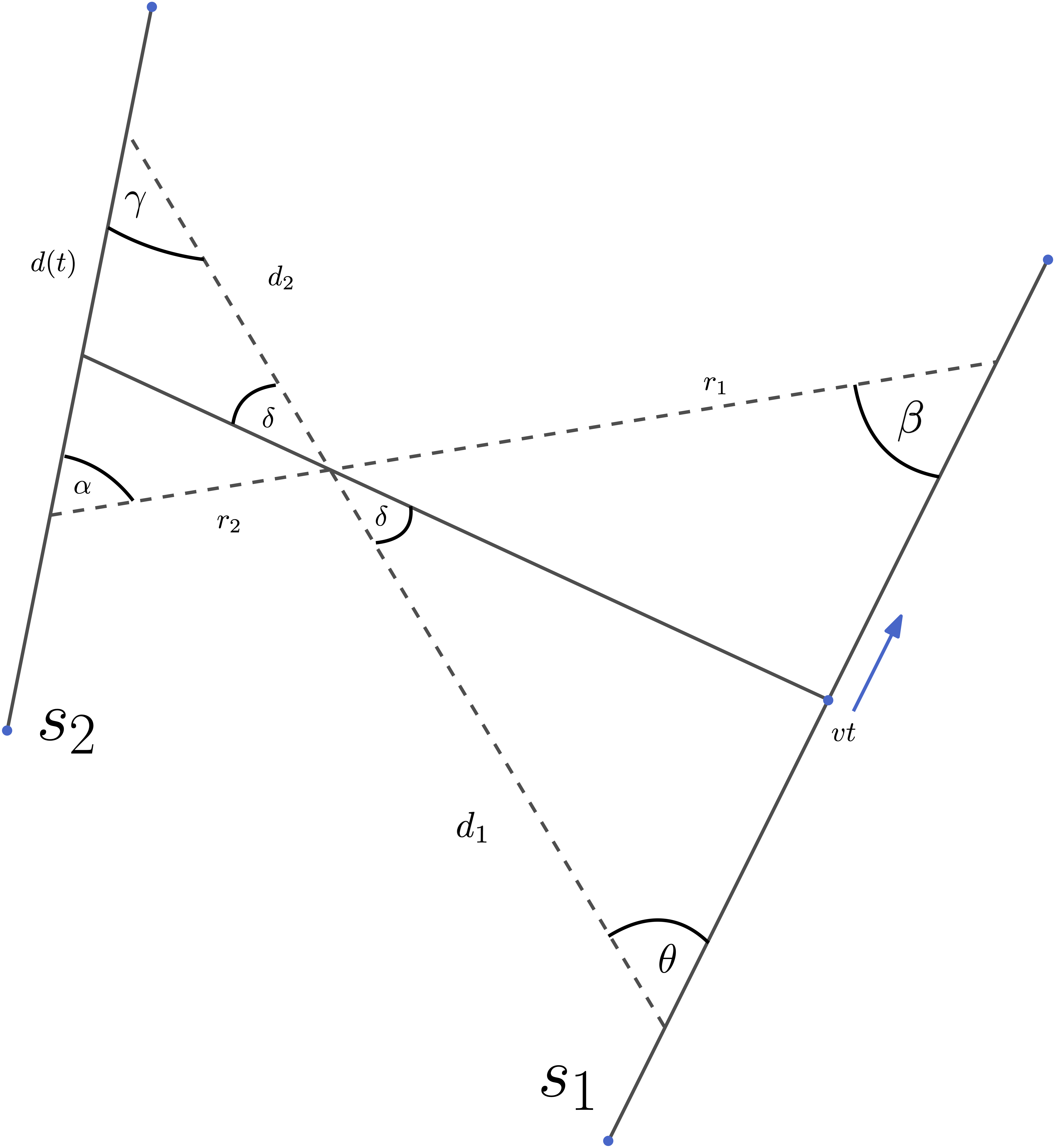}
    \caption{An entity travels along segment $s_1$ at constant speed $v$, covering distance $vt$ from its start. The ray from this entity to segment $s_2$ meets the bi-tangent of $L(s_1, s_2)$ at angle $\delta$, while the fixed angles between the bi-tangent and the upper and lower chains are $\theta$ and $\gamma$, respectively. Distances $d_1$ and $d_2$ are measured from the endpoints of $s_1$ and $s_2$ to the corresponding tangency points.}
    \label{fig:MovingEntity}
\end{figure}

Observe that the mapping from positions on $\tau_q$ to positions on $\tau_r$ is non-linear. For example, an entity moving along a segment $s_1$ at constant speed $v$ maps to a point on another segment $s_2$ via a nonlinear function $f(t)$, where the distance $d(t)$ from a fixed endpoint of $s_2$ is strictly convex (or concave).

\begin{observation}
\label{obs::mapOnOtherTraj}
Let $s_1, s_2 \subset P$ be two line segments. Suppose an entity moves with constant speed $v$ on $s_1$. Construct $L(s_1, s_2)$ and consider its intersection points. Define a mapping of the entity's position as $f(t)$ to be a point on $s_2$ at distance $d(t)$ from a certain endpoint of $s_2$ at time $t$. Then, as the entity moves, the mapping $f(t)$ transitions between the intersection points, and $d(t)$ has second derivative of constant sign (strictly convex or concave, or linear when $\theta = \gamma$) in $t$.
\end{observation}

\begin{proof}
Let $d(t)$ be denoted as $x$. Referring to Figure~\ref{fig:MovingEntity} and applying the sine law on the two relevant triangles, we obtain:
\[
v t = \frac{d_1 \sin\delta}{\sin(\delta + \theta)}
\quad\text{and}\quad
x = \frac{d_2 \sin\delta}{\sin(\gamma + \delta)}
\]
Eliminating $\delta$ gives the closed form:
\[
x = \frac{d_2 v t \sin\theta}{d_1 \sin\gamma + v t \sin(\theta - \gamma)}
\]
Now set:
\[
A = \frac{\sin(\theta - \gamma)}{\sin\theta}, \qquad B = \frac{d_1 \sin\gamma}{v \sin\theta}
\]
so that $x = \frac{d_2 t}{A t + B} = d_2 \cdot \frac{t}{B} - d_2 \cdot \frac{A t^2}{B(A t + B)}$. If $\theta = \gamma$, then $A = 0$ and $x$ is linear in $t$. Otherwise, a straightforward differentiation gives $d''(t) = -2 d_2 A B / (A t + B)^3$, which keeps a constant non-zero sign throughout the valid subsegment (under non-degeneracy conditions, where $A t + B > 0$ and the angles are non-zero).
\end{proof}

Consider any subsegment of $\tau_q$ between two consecutive points of $I_q$ (and similarly of $\tau_r$ between two consecutive points of $I_r$). Observe that along such a subsegment, no additional intersection with a critical constraint is encountered on either trajectory; hence the same two reflex vertices determine the first/last mutually visible points on the opposite trajectory throughout this interval. Therefore, over this interval, the boundary of the mutual-visibility region in $D$ is traced by a mapping of the type in Observation~\ref{obs::mapOnOtherTraj}. Moreover, by Observation~\ref{obs::mapOnOtherTraj}, each corresponding boundary portion in $D$ has constant curvature sign (or is a straight segment) between its two endpoints; in particular, it lies entirely on one side of its chord, so the region enclosed by this curve and the line segment connecting its endpoints is convex.

Given query speeds $v_q$ and $v_r$ for $q$ and $r$ during a constant-speed phase starting at $(x_0, y_0) \in D$ (with $(x_0, y_0) = (0, 0)$ for $[0, \tmin]$ and $(x_0, y_0) \in \partial D$ for $[\tmin, t_f]$), their positions trace the ray $\overrightarrow{r}$ from the phase's starting point $(x_0, y_0)$ in direction $(v_q, v_r)$ (allowing $v_q = 0$ or $v_r = 0$). Each arc endpoint on $\partial V$ coincides with a critical-constraint intersection with the trajectories, and there are $\mathcal{O}(n)$ such intersections overall; hence $\partial V$ is a closed chain of $\mathcal{O}(n)$ piecewise-algebraic arcs, so $V$ is a \emph{splinegon}. Cutting $D \setminus V$ along the two vertical segments through the leftmost and rightmost points of $V$ to $\partial D$ partitions $D \setminus V$ into at most six pairwise interior-disjoint simple splinegons of total complexity $\mathcal{O}(n)$ (the left and right rectangles, together with at most two regions above $V$ and at most two below $V$, since by simplicity of $P$ the region $V$ meets each horizontal edge of $D$ in at most one interval). We preprocess each of these simple splinegons in $\mathcal{O}(n)$ time and space for $\mathcal{O}(\log n)$-time splinegon ray shooting~\cite{melissaratos1992shortest,GHODSI2014435}. Since $V$ is $x$-monotone, a binary search on the sorted $x$-coordinates of the vertices of $\partial V$ locates the unique upper and lower arcs of $\partial V$ at $x_0$ in $\mathcal{O}(\log n)$ time, determining whether $(x_0, y_0) \in V$ (in which case visibility holds at the start of the phase) or which simple splinegon component of $D \setminus V$ contains $(x_0, y_0)$. In the latter case, tracing $\overrightarrow{r}$ across at most two vertical cuts until it first meets $\partial V$ (or exits $D$) reports $t^*$ in $\mathcal{O}(\log n)$ time. Therefore, the following theorem holds:

\begin{theorem}
\label{theorem::FixedTrajectories}
Constructing $D$ in $\mathcal{O}(n)$ time, using $\mathcal{O}(n)$ space, one can process any query $(v_q,v_r)$ to determine in $\mathcal{O}(\log n)$ time the smallest $t^*\ge0$ for which $q$ and $r$ become mutually visible.
\end{theorem}


\section{One Query-Provided Trajectory}
\label{sec::OneQueryProvidedTrajectory}

In this section, the trajectory $\tau_r \subset P$ of entity $r$ is fixed in advance (symmetrically if $\tau_q$ is fixed), while the trajectory $\tau_q \subset P$ of entity $q$ and both speeds $v_q, v_r$ are provided at query time.

\medskip
\noindent\textbf{Preprocessing.} We first compute the weak visibility polygon $W(\tau_r)$ in $\mathcal{O}(n)$ time~\cite{ghosh2007}. Since any line of sight between $\tau_q$ and $\tau_r$ lies entirely within $W(\tau_r)$, we preprocess $W(\tau_r)$ for $\mathcal{O}(\log n)$-time ray shooting~\cite{chazelle1994ray}, and triangulate and preprocess the planar subdivision of $P$ formed by $W(\tau_r)$ and the pockets of $P \setminus W(\tau_r)$ for $\mathcal{O}(\log n)$-time point location~\cite{chazelle1991triangulating,ghosh2007,kirkpatrick1983optimal}, using $\mathcal{O}(n)$ time and space.

Next, we construct the shortest-path trees $T(p^+_{\tau_r})$ and $T(p^-_{\tau_r})$ within $W(\tau_r)$, along with their triangulated shortest-path maps for $\mathcal{O}(\log n)$-time point location, in $\mathcal{O}(n)$ time and space~\cite{chazelle1991triangulating,guibas1987linear,kirkpatrick1983optimal}. For every vertex $v$ in either tree, we also store its depth and a pointer $\mathrm{up}_j(v)$ to its $2^j$-th ancestor for each $0 \le j \le \lfloor \log_2 \operatorname{depth}(v) \rfloor$. This requires $\mathcal{O}(n \log n)$ time and space and allows us to access any vertex along a root-to-vertex path in $\mathcal{O}(\log n)$ time.

Finally, we decompose $T(p^+_{\tau_r})$ and $T(p^-_{\tau_r})$ into vertex-disjoint solid paths stored as biased binary trees~\cite{sleator1983dynamic}. Choosing a coordinate system in which the supporting line of $\tau_r$ is the $x$-axis, every tree edge $e = \overline{w_1 w_2}$ not parallel to $\tau_r$ has supporting line $x + s_e y - d_e = 0$, where
\[
s_e = -\frac{x_{w_2} - x_{w_1}}{y_{w_2} - y_{w_1}} \quad\text{and}\quad d_e = x_{w_1} + s_e y_{w_1}.
\]
We lift each such edge $e$ to the point $p(e) = (s_e, d_e, s_e d_e) \in \mathbb{R}^3$; edges parallel to $\tau_r$ are omitted, as each query splits $\tau_q$ where it crosses the supporting line of $\tau_r$ (so the line of sight is never parallel to $\tau_r$ on open subintervals). For each node $\nu$ of every biased binary tree covering $m_\nu$ lifted edges, we construct the 3D convex hull of its lifted points bottom-up in $\mathcal{O}(m_\nu)$ time using linear-time 3D hull merging~\cite{chazelle1992optimal}, and store the hull in a Dobkin--Kirkpatrick hierarchy~\cite{dobkin1990determining} of size $\mathcal{O}(m_\nu)$. Since each solid edge lies at depth $\mathcal{O}(\log n)$ in its biased binary tree~\cite{sleator1983dynamic} (while off-path edges are stored separately), we have $\sum_\nu m_\nu = \mathcal{O}(n \log n)$. Thus, the overall preprocessing takes $\mathcal{O}(n \log n)$ time and $\mathcal{O}(n \log n)$ space.

\begin{figure}[t]
     \centering
     \begin{subfigure}[b]{0.45\textwidth}
         \centering
         \includegraphics[width=\textwidth]{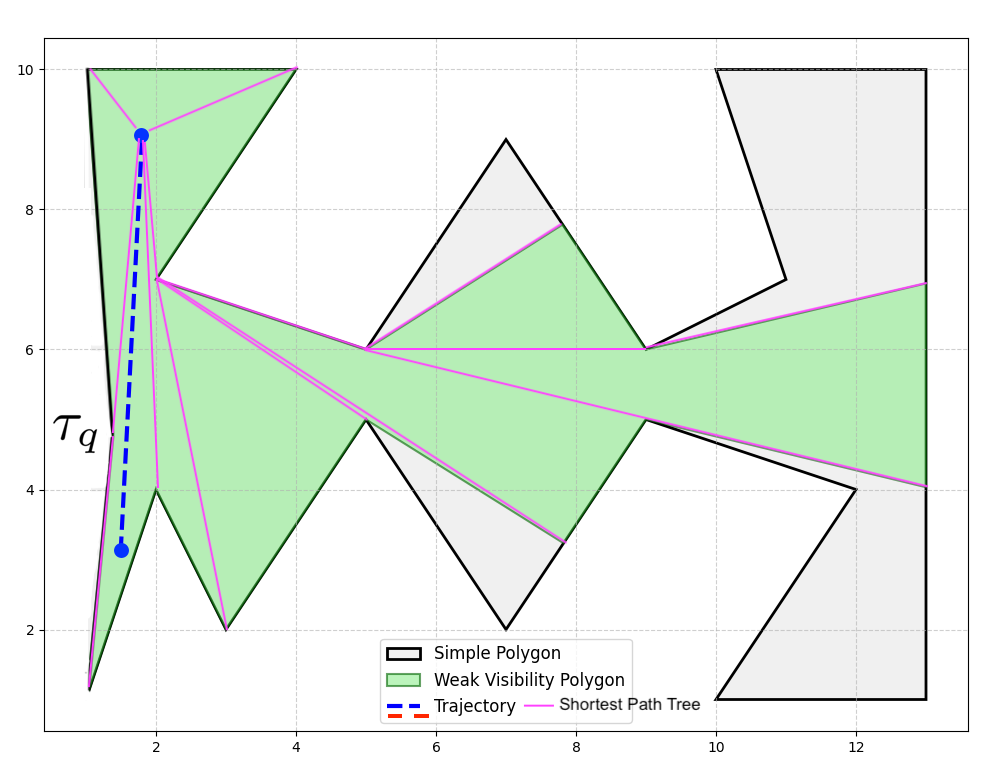}
         \caption{Fixed trajectory (blue, illustrated with $\tau_q$ fixed) together with its shortest-path tree (purple) within its weak visibility polygon.}
         \label{fig:spt}
     \end{subfigure}
     \hfill
     \begin{subfigure}[b]{0.48\textwidth}
         \centering
         \includegraphics[width=\textwidth]{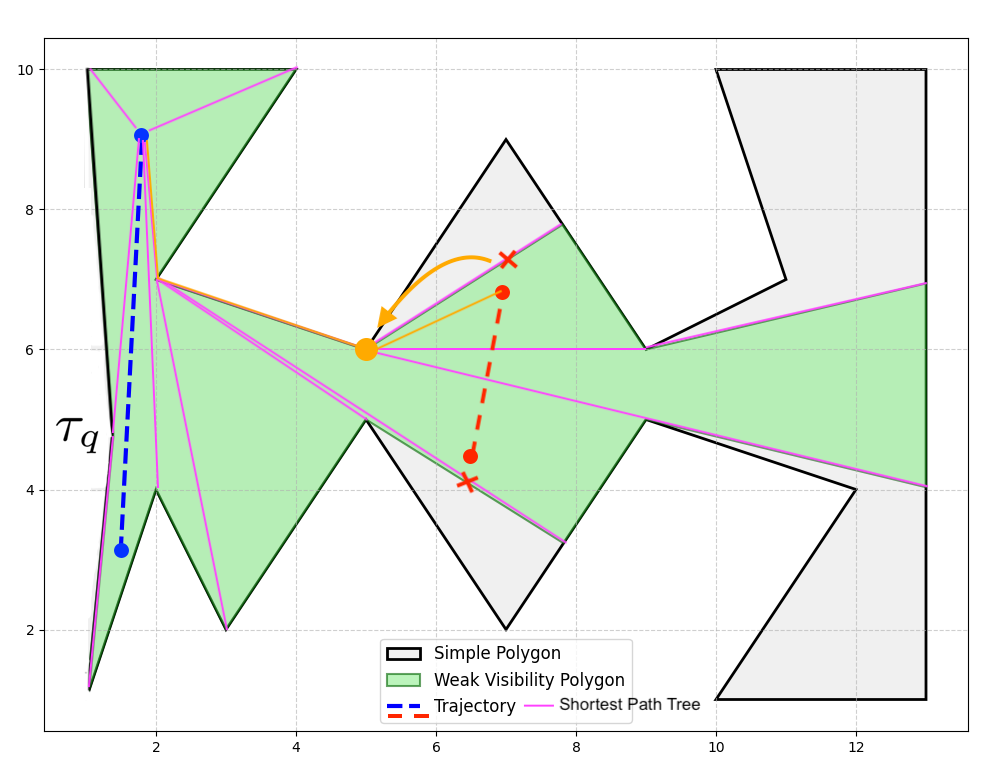}
         \caption{Identifying the visibility glass for the query trajectory.}
         \label{fig:vg}
     \end{subfigure}
        \caption{Finding the visibility glass when one trajectory is given as a query (illustrated symmetrically with $\tau_q$ fixed in blue). (a) The shortest-path tree (purple) is precomputed within the weak visibility polygon. (b) The precomputed trees determine the boundary chains of the visibility glass.}
        \label{fig:sptVG}
\end{figure}

\medskip
\noindent\textbf{Finding the visibility glass.} Upon receiving $\tau_q, v_q,$ and $v_r$, we first restrict $\tau_q$ to its intersection with $W(\tau_r)$ in $\mathcal{O}(\log n)$ time, discarding the remainder. Locating the two endpoints of $\tau_q$ in the precomputed planar subdivision of $P$ takes $\mathcal{O}(\log n)$ time. Because $\tau_q \subset P$ is a line segment and cannot cross $\partial P$, if both endpoints lie in the same pocket of $P \setminus W(\tau_r)$, then $\tau_q \cap W(\tau_r) = \emptyset$ and no visibility occurs. Otherwise, $\tau_q \cap W(\tau_r)$ is a single contiguous segment whose endpoints are obtained by intersecting $\tau_q$ with the unique window bounding each endpoint's pocket (if that endpoint lies outside $W(\tau_r)$). Restricting the active time interval to the subinterval during which $q(t) \in \tau_q \cap W(\tau_r)$ and denoting $\tau_q \cap W(\tau_r)$ simply by $\tau_q$, every ray-shooting test below runs inside $W(\tau_r)$. 

Let $L_q$ and $L_r$ denote the supporting lines of $\tau_q$ and $\tau_r$. If $L_q = L_r$ (the collinear case), two $\mathcal{O}(\log n)$-time ray-shooting queries inside $W(\tau_r)$ from an endpoint of $\tau_r$ along both directions of $L_r$ find the maximal chord $I_0 \subseteq W(\tau_r) \cap L_r$ containing $\tau_r$; then $\overline{q(t)r(t)} \subseteq P$ holds iff $q(t) \in I_0$, which yields $t^*$ in $\mathcal{O}(1)$ time. Otherwise ($L_q \neq L_r$), we split the active time interval at the (at most two) instants where $q(t) \in L_r$ or $r(t) \in L_q$. This yields at most three subintervals processed in chronological order. On each subinterval, the active subsegments $s_q \subseteq \tau_q$ and $s_r \subseteq \tau_r$ lie in closed halfplanes of $L_r$ and $L_q$, respectively, so their endpoints lie in convex position (or share a single endpoint) and $y_q(t)$ maintains a constant strict sign on the open subinterval.

Locating $p^+_{s_q}$ and $p^-_{s_q}$ in the shortest-path maps of $T(p^+_{\tau_r})$ and $T(p^-_{\tau_r})$ in $\mathcal{O}(\log n)$ time gives their parent vertices in both trees (Figure~\ref{fig:sptVG}). For any endpoint $a \in \{p^+_{s_q}, p^-_{s_q}\} \subset W(\tau_r)$ and any endpoint $b \in \{p^+_{s_r}, p^-_{s_r}\} \subseteq \tau_r$, since $a \in W(\tau_r)$, $a$ sees at least one point on $\tau_r$, so $\pi(a, p^+_{\tau_r})$ and $\pi(a, p^-_{\tau_r})$ are already inward-convex chains forming the boundary of the funnel $\mathcal{F}(a, \tau_r)$ with apex $a$. Hence $\pi(a, b)$ is either the segment $\overline{ab}$ (checked by $\mathcal{O}(\log n)$ ray shooting) or a prefix of $\pi(a, p^+_{\tau_r})$ or $\pi(a, p^-_{\tau_r})$ up to the unique tangent vertex $z$ from $b$ (located in $\mathcal{O}(\log^2 n)$ time by binary search using $\mathrm{up}_j$), followed by $\overline{zb}$. Thus all four endpoint-to-endpoint paths $\pi(p^+_{s_q}, p^+_{s_r})$, $\pi(p^-_{s_q}, p^-_{s_r})$, $\pi(p^+_{s_r}, p^-_{s_q})$, and $\pi(p^-_{s_r}, p^+_{s_q})$ are represented as contiguous subpaths in $T(p^+_{\tau_r})$ or $T(p^-_{\tau_r})$ plus at most two terminal edges, supporting $\mathcal{O}(\log n)$-time vertex access via $\mathrm{up}_j$.

If $L(s_q, s_r)$ is non-empty, each cross-path $\pi(p^+_{s_r}, p^-_{s_q})$ and $\pi(p^-_{s_r}, p^+_{s_q})$ switches turn direction between clockwise and counterclockwise at most once (Observation~1 of~\cite{swat2020}). For each cross-path: if the endpoints are mutually visible, we take that single segment; if all turns along the path are in the same sense (zero inflection edges), the inner bi-tangent is whichever of its two terminal edges (incident to its two endpoints) extends inside $W(\tau_r)$ to intersect the opposite segment (verified in $\mathcal{O}(\log n)$ time by ray shooting); otherwise, we binary-search the path in $\mathcal{O}(\log^2 n)$ time using $\mathrm{up}_j$ for its unique inflection edge. Extending these two inner bi-tangents to meet $s_q$ and $s_r$ (verified by $\mathcal{O}(\log n)$ ray shooting; if either fails to cut $s_q$ and $s_r$ inside $W(\tau_r)$, then $L(s_q, s_r) = \emptyset$ and we skip this subinterval) yields subsegments $q' \subseteq s_q$ and $r' \subseteq s_r$ bounding the visibility glass $L(s_q, s_r)$ (if the two bi-tangents coincide, $L(s_q, s_r)$ is a single segment and visibility is checked in $\mathcal{O}(\log n)$ time). Because $L(s_q, s_r)$ is bounded by these inner bi-tangents and $s_q, s_r$ lie in convex position, its upper and lower chains $\gamma^+ = \pi(p^+_{q'}, p^+_{r'})$ and $\gamma^- = \pi(p^-_{q'}, p^-_{r'})$ are inward-convex (meeting at most along a single segment), the interior vertices and edges of $\gamma^+$ form a contiguous subpath of $\pi(p^+_{s_q}, p^+_{s_r})$ in whichever of $T(p^+_{\tau_r})$ or $T(p^-_{\tau_r})$ contains $\pi(p^+_{s_q}, p^+_{s_r})$, and those of $\gamma^-$ form a contiguous subpath of $\pi(p^-_{s_q}, p^-_{s_r})$ in whichever of the two trees contains $\pi(p^-_{s_q}, p^-_{s_r})$. Restricting the current time subinterval to $[t'_0, t'_1]$ where $q(t) \in q'$ and $r(t) \in r'$ (terminating if empty), we test $t'_0$ by an $\mathcal{O}(\log n)$ ray-shooting query; if $t'_0$ is visible, we return $t^* = t'_0$.

\medskip
\noindent\textbf{Visibility condition.}
For any point $v = (x_v, y_v)$, define
\begin{equation}
\label{eq:fv}
f_v(t) := (x_q(t) - x_v)(y_r(t) - y_v) - (y_q(t) - y_v)(x_r(t) - x_v),
\end{equation}
which equals $(x_r(t) - x_q(t))(y_v - y_q(t)) - (y_r(t) - y_q(t))(x_v - x_q(t))$. Thus $f_v(t) \ge 0$ if and only if $v$ lies on or to the left of the directed line $g(t) = \overrightarrow{q(t)r(t)}$, and $f_v(t) \le 0$ if and only if $v$ lies on or to the right of $g(t)$. Labeling $\gamma^+$ as the chain on the left side of $L(s_q, s_r)$ (viewed from $q'$ toward $r'$) and $\gamma^-$ as the chain on the right side, and defining
\[
G^+(t) := \min_{u \in \gamma^+} f_u(t)
\quad\text{and}\quad
G^-(t) := \max_{w \in \gamma^-} f_w(t),
\]
we have for every $t \in [t'_0, t'_1]$:
\begin{equation}
\label{eq:visCond}
\overline{q(t)r(t)} \subseteq P \iff G^+(t) \ge 0 \;\text{ and }\; G^-(t) \le 0.
\end{equation}

\begin{observation}
\label{obs::collinearityCandidate}
For every vertex $v = (x_v, y_v)$, $f_v(t) = A t^2 + B_v t + C_v$ is a quadratic polynomial whose leading coefficient $A = v_{qx} v_{ry} - v_{qy} v_{rx}$ is independent of $v$. Consequently, any transition into mutual visibility on $(t'_0, t'_1]$ occurs at a non-negative real root of $f_v(t) = 0$ for some vertex $v \in \gamma^+ \cup \gamma^-$ such that $\overline{q(t)r(t)} \subseteq P$.
\end{observation}

\begin{proof}
Setting $\Delta x_{q0} = x_{q,0} - x_v$, $\Delta y_{q0} = y_{q,0} - y_v$, $\Delta x_{r0} = x_{r,0} - x_v$, and $\Delta y_{r0} = y_{r,0} - y_v$, expanding~\eqref{eq:fv} gives $f_v(t) = A t^2 + B_v t + C_v$ with
\begin{align*}
A &= v_{qx} v_{ry} - v_{qy} v_{rx}, \\
B_v &= \Delta x_{q0} v_{ry} - \Delta y_{q0} v_{rx} + \Delta y_{r0} v_{qx} - \Delta x_{r0} v_{qy}, \\
C_v &= \Delta x_{q0} \Delta y_{r0} - \Delta y_{q0} \Delta x_{r0}.
\end{align*}
In particular, $A$ depends only on the speed vectors of $q$ and $r$. If $A \neq 0$, the equation $f_v(t) = 0$ has at most two real roots; if $A = 0$ and $B_v \neq 0$, it has the unique root $t = -C_v / B_v$; if $A = B_v = 0$, $v$ is either collinear with $q(t)$ and $r(t)$ for all $t$ ($C_v = 0$) or for no $t$ ($C_v \neq 0$). Since visibility does not hold at $t'_0$, continuity of $G^+$ and $G^-$ implies that at the earliest $t^* \in (t'_0, t'_1]$ satisfying~\eqref{eq:visCond}, either $G^+(t^*) = 0$ or $G^-(t^*) = 0$, meaning $f_v(t^*) = 0$ for some $v \in \gamma^+ \cup \gamma^-$.
\end{proof}

\begin{figure}[t]
     \centering
     \begin{subfigure}[b]{0.48\textwidth}
         \centering
         \includegraphics[width=\textwidth]{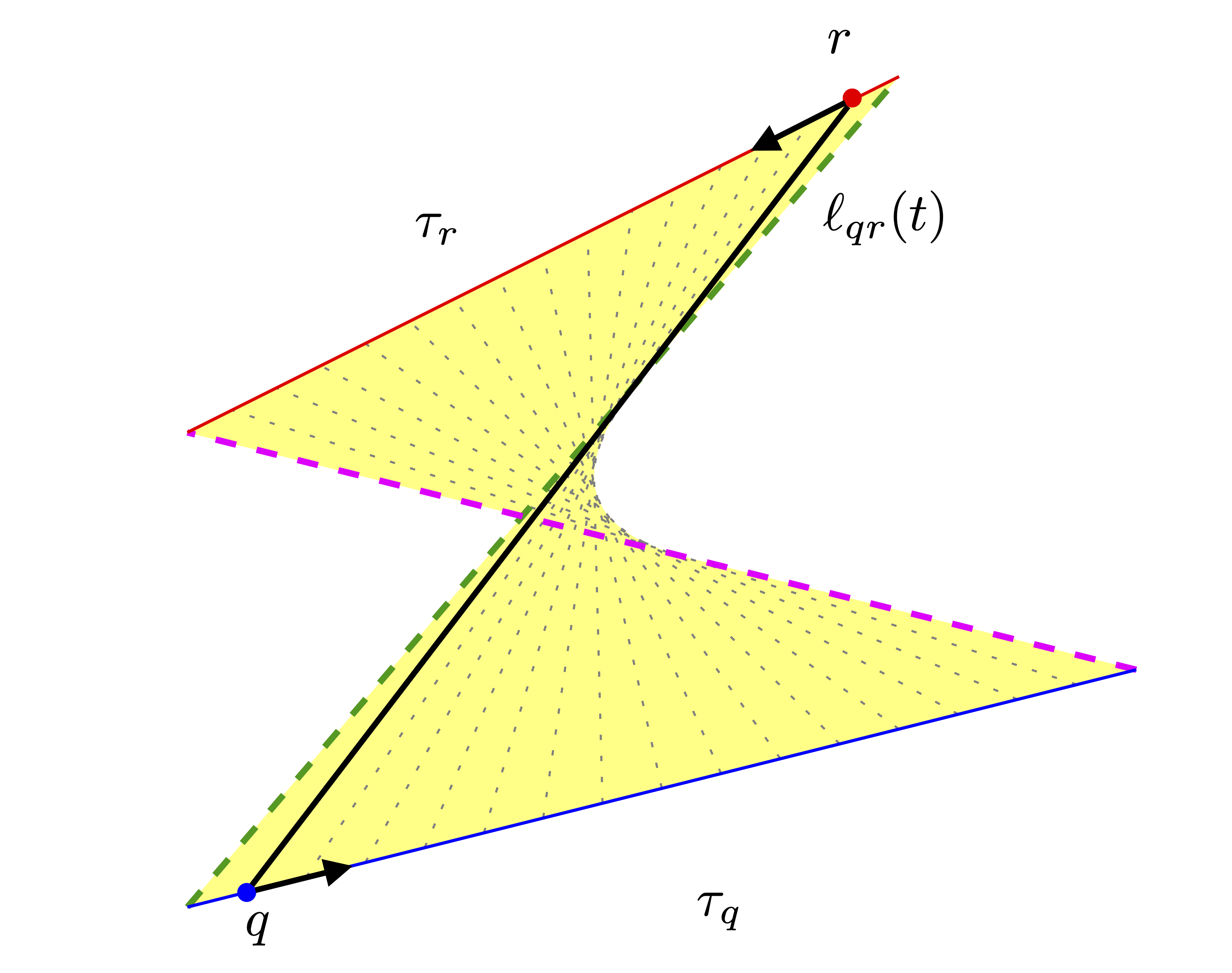}
         \caption{Opposite direction (with a degree-two envelope).}
         \label{fig:sweptOpp}
     \end{subfigure}
     \hfill
     \begin{subfigure}[b]{0.45\textwidth}
         \centering
         \includegraphics[width=\textwidth]{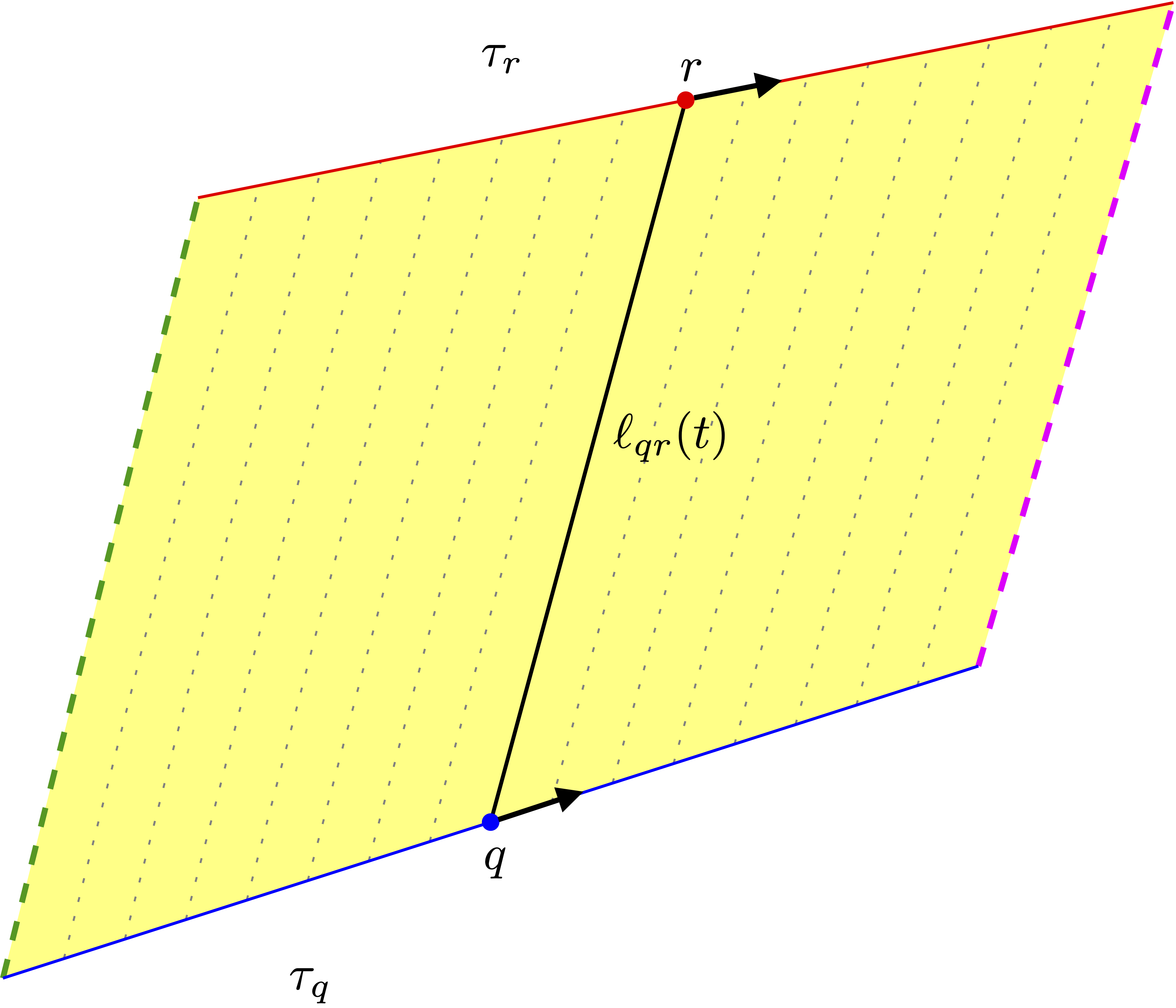}
         \caption{Similar direction.}
         \label{fig:sweptSame}
     \end{subfigure}
        \caption{The area $S$ (yellow) swept by the leash $\ell_{qr}(t)$ between $q$ ($\tau_q$, blue) and $r$ ($\tau_r$, red). In opposite-direction motion with non-parallel speed vectors (a), the moving line $g(t)$ wraps around a degree-two envelope, which allows the line of sight to clear and re-hit one of the chains of $L(\tau_q, \tau_r)$ multiple times.}
        \label{fig:figTrajs}
\end{figure}

Consider the moving \emph{leash} $\ell_{qr}(t) = \overline{q(t)r(t)}$ and the area $S = \bigcup_{t \in [0, t_f]} \ell_{qr}(t)$ it sweeps (Figure~\ref{fig:figTrajs}). Viewing~\eqref{eq:fv} as an equation in $(x, y) = (x_v, y_v)$ writes the moving line $g(t) = \overleftrightarrow{q(t)r(t)}$ as $A t^2 + B(x, y) t + C(x, y) = 0$, where $B(x, y)$ and $C(x, y)$ are linear in $(x, y)$. When $A \neq 0$, eliminating $t$ between this quadratic and its $t$-derivative shows that the lines $\{g(t)\}$ wrap around the degree-two envelope $B(x, y)^2 - 4 A C(x, y) = 0$. Consequently, when $A < 0$, every $f_u(t)$ is a downward-opening quadratic ($f_u''(t) = 2A < 0$), so $G^+(t) = \min_{u \in \gamma^+} f_u(t)$ is concave (as the pointwise minimum of concave quadratics), making $\{t \in [t'_0, t'_1] \mid G^+(t) \ge 0\}$ a single interval. By contrast, $G^-(t) = \max_{w \in \gamma^-} f_w(t)$ is the pointwise maximum of downward-opening quadratics: as $g(t)$ rolls around this degree-two envelope, $G^-(t)$ arches downward between consecutive edge-parallel instants and can dip below $0$ on arbitrarily many disconnected intervals (Figure~\ref{fig:sweptOpp}).

Express all coordinates in the frame where $\tau_r$ lies on the $x$-axis ($y_{r,0} = v_{ry} = 0$) and assume without loss of generality that $y_q(t) > 0$ on $(t'_0, t'_1)$ (the case $y_q(t) < 0$ is symmetric). Set
\[
N_0 := (v_{rx} - v_{qx}) y_{q,0} - (x_{r,0} - x_{q,0}) v_{qy}.
\]
The direction angle $\theta(t) \in (-\pi, 0)$ of $g(t)$ from $q(t)$ to $r(t)$ has derivative
\[
\theta'(t) = \frac{\bigl(x_r(t)-x_q(t)\bigr)(-v_{qy}) - \bigl(-y_q(t)\bigr)(v_{rx}-v_{qx})}{\|r(t) - q(t)\|^2} = \frac{N_0}{\|r(t) - q(t)\|^2}.
\]

\medskip
\noindent\textbf{Constant-direction case ($N_0 = 0$).}
When $N_0 = 0$, we have $\theta'(t) = 0$ for all $t \in (t'_0, t'_1)$, so the directed line $g(t)$ keeps a constant direction throughout $(t'_0, t'_1)$ (as $y_q(t) > 0$ prevents $r(t) - q(t)$ from vanishing or reversing). A single binary search along $\gamma^+$ (and along $\gamma^-$) using the $\mathrm{up}_j$ pointers locates in $\mathcal{O}(\log^2 n)$ time an extreme vertex $u^* \in \gamma^+$ attaining $G^+(t) = f_{u^*}(t)$, and an extreme vertex $w^* \in \gamma^-$ attaining $G^-(t) = f_{w^*}(t)$ for all $t \in [t'_0, t'_1]$. Solving the two constant-degree inequalities $f_{u^*}(t) \ge 0$ and $f_{w^*}(t) \le 0$ on $[t'_0, t'_1]$ takes $\mathcal{O}(1)$ time and yields $t^*$.

\medskip
\noindent\textbf{The visible interval for $\gamma^+$ ($A \le 0, N_0 \neq 0$).}
When $N_0 \neq 0$, the direction angle $\theta(t)$ of $g(t)$ rotates monotonically over $(t'_0, t'_1)$. Since $\gamma^+ = (u_0, u_1, \dots, u_k, u_{k+1})$ (with $u_0 = p^+_{q'}$ and $u_{k+1} = p^+_{r'}$) is inward-convex, its edges also turn monotonically, so the vertex of $\gamma^+$ attaining $G^+(t) = \min_{u \in \gamma^+} f_u(t)$ advances monotonically along a contiguous subchain $u_a, \dots, u_b$ of $\gamma^+$. Specifically, this vertex switches from $u_j$ to $u_{j+1}$ at the unique time $t_j \in (t'_0, t'_1)$ when $g(t)$ becomes parallel to $\overline{u_j u_{j+1}}$, which is computed in $\mathcal{O}(1)$ time. We locate $u_a$ and $u_b$ in $\mathcal{O}(\log^2 n)$ time by two binary searches along $\gamma^+$ (using the $\mathrm{up}_j$ pointers) matching the directions $\theta(t'_0)$ and $\theta(t'_1)$. Indexing this subchain in increasing order of time (reversing it if needed) and setting $t_{a-1} = t'_0$ and $t_b = t'_1$, we obtain $t_{a-1} \le t_a < \dots < t_{b-1} \le t_b$ such that $G^+(t) = f_{u_j}(t)$ on each interval $[t_{j-1}, t_j]$ ($a \le j \le b$).

Whenever $A \le 0$, $G^+(t)$ has non-increasing one-sided derivatives on $[t'_0, t'_1]$, so $\{t \in [t'_0, t'_1] \mid G^+(t) \ge 0\}$ is a single (possibly empty) closed interval $[t_{\mathrm{lo}}, t_{\mathrm{hi}}]$. We find $[t_{\mathrm{lo}}, t_{\mathrm{hi}}]$ in $\mathcal{O}(\log^2 n)$ time as follows:
\begin{enumerate}
    \item \emph{Finding the peak $t_{\mathrm{peak}}$:} Binary-search $j \in \{a, \dots, b\}$ using the $\mathrm{up}_j$ pointers ($\mathcal{O}(\log n)$ time per step). At index $j$, compute $f'_{u_j}(t_{j-1})$ and $f'_{u_j}(t_j)$ in $\mathcal{O}(1)$ time. Since the slope of $G^+$ is non-increasing, if $f'_{u_j}(t_{j-1}) \le 0$, then $t_{\mathrm{peak}} \le t_{j-1}$ (recurse left); if $f'_{u_j}(t_j) \ge 0$, then $t_{\mathrm{peak}} \ge t_j$ (recurse right); otherwise $t_{\mathrm{peak}}$ lies inside $(t_{j-1}, t_j)$ at the unique zero of $f'_{u_j}(t) = 0$ (with $p = j$). At the boundaries, if $f'_{u_a}(t'_0) \le 0$ then $t_{\mathrm{peak}} = t'_0$ ($p = a$); if $f'_{u_b}(t'_1) \ge 0$ then $t_{\mathrm{peak}} = t'_1$ ($p = b$); and if the search converges between adjacent indices $j$ and $j + 1$, then $t_{\mathrm{peak}} = t_j$ ($p = j + 1$). If $G^+(t_{\mathrm{peak}}) < 0$, the clearing interval is empty and we terminate.
    \item \emph{Finding $t_{\mathrm{lo}}$ and $t_{\mathrm{hi}}$:} Because $G^+$ is non-decreasing on $[t'_0, t_{\mathrm{peak}}]$, if $G^+(t'_0) \ge 0$ then $t_{\mathrm{lo}} = t'_0$; otherwise, if $G^+(t_{p-1}) \le 0$, then $t_{\mathrm{lo}}$ is the zero of $f_{u_p}(t) = 0$ on $[t_{p-1}, t_{\mathrm{peak}}]$. Otherwise ($G^+(t_{p-1}) > 0$), binary-searching $j \in \{a, \dots, p - 1\}$ by the sign of $G^+(t_j) = f_{u_j}(t_j)$ in $\mathcal{O}(\log^2 n)$ time (every such $t_j$ lies in $[t'_0, t_{\mathrm{peak}}]$) locates the unique active vertex $u_l$ across which $G^+$ crosses $0$, and solving $f_{u_l}(t) = 0$ on $[t_{l-1}, t_l]$ in $\mathcal{O}(1)$ time gives $t_{\mathrm{lo}}$. Symmetrically, on $[t_{\mathrm{peak}}, t'_1]$ (checking $G^+(t'_1) \ge 0$ and $G^+(t_p) \le 0$, and otherwise binary-searching $j \in \{p + 1, \dots, b\}$ by the sign of $G^+(t_{j-1}) = f_{u_j}(t_{j-1})$), we obtain $t_{\mathrm{hi}}$.
\end{enumerate}

\medskip
\noindent\textbf{Parallel speed vectors ($A = 0, N_0 \neq 0$).}
When $A = 0$, every $f_v(t)$ is linear in $t$, so both $G^+(t)$ and $-G^-(t)$ have non-increasing one-sided derivatives on $[t'_0, t'_1]$. Running the $\mathcal{O}(\log^2 n)$-time binary search above on $\gamma^+$ and on $\gamma^-$ gives the two clearing intervals $[t^+_{\mathrm{lo}}, t^+_{\mathrm{hi}}]$ and $[t^-_{\mathrm{lo}}, t^-_{\mathrm{hi}}]$; if their intersection is non-empty, its left endpoint is $t^*$.

\medskip
\noindent\textbf{Non-parallel speed vectors ($A \neq 0, N_0 \neq 0$).}
Assume $A < 0$ (the case $A > 0$ is symmetric, exchanging the roles of $\gamma^+$ and $\gamma^-$ and negating signs). We first compute the single clearing interval $[t_{\mathrm{lo}}, t_{\mathrm{hi}}] \subseteq [t'_0, t'_1]$ of $\gamma^+$ in $\mathcal{O}(\log^2 n)$ time as above (terminating if empty). It remains to find the smallest $t \in [t_{\mathrm{lo}}, t_{\mathrm{hi}}]$ with $G^-(t) \le 0$.

First, test $t_{\mathrm{lo}}$ by an $\mathcal{O}(\log n)$ ray-shooting query; if $\overline{q(t_{\mathrm{lo}})r(t_{\mathrm{lo}})} \subseteq P$, return $t^* = t_{\mathrm{lo}}$. Otherwise, $G^-(t_{\mathrm{lo}}) > 0$. If $\gamma^-$ has no interior reflex vertices, then $G^-(t) \le 0$ holds identically on $[t_{\mathrm{lo}}, t_{\mathrm{hi}}]$, which would already have returned $t_{\mathrm{lo}}$. Otherwise, let $e = \overline{w_j w_{j+1}}$ be any edge of $\gamma^-$ whose parallel instant $t_e$ lies in $(t_{\mathrm{lo}}, t_{\mathrm{hi}})$. Between any two consecutive instants among $\{t_{\mathrm{lo}}, t_{\mathrm{hi}}\} \cup \{t_e \in (t_{\mathrm{lo}}, t_{\mathrm{hi}})\}$, a single vertex $w \in \gamma^-$ is active, so $G^-(t) = f_w(t)$ is a downward-opening quadratic and attains its minimum on that subinterval at an endpoint. Therefore, mutual visibility occurs on $[t_{\mathrm{lo}}, t_{\mathrm{hi}}]$ if and only if either $f_{w_j}(t_e) \le 0$ for some edge $e = \overline{w_j w_{j+1}} \in \gamma^-$ with $t_e \in (t_{\mathrm{lo}}, t_{\mathrm{hi}})$, or $G^-(t_{\mathrm{hi}}) \le 0$.

Since $y_q(t) > 0$ on $(t_{\mathrm{lo}}, t_{\mathrm{hi}}) \subseteq (t'_0, t'_1)$, the line $g(t)$ is never horizontal on $(t_{\mathrm{lo}}, t_{\mathrm{hi}})$. Consider any non-horizontal edge $e = \overline{w_j w_{j+1}}$ of $\gamma^-$, whose supporting line is $x + s_e y - d_e = 0$. At the time $t_e$ when $g(t_e)$ becomes parallel to $e$, its two points $q(t_e) = (x_q(t_e), y_q(t_e))$ and $r(t_e) = (x_r(t_e), 0)$ share the same value of $x + s_e y$, so $x_q(t_e) + s_e y_q(t_e) = x_r(t_e)$. Solving for $t_e$ gives
\[
t_e = -\frac{(x_{q,0} - x_{r,0}) + s_e y_{q,0}}{D(s_e)},
\qquad
D(s_e) := (v_{qx} - v_{rx}) + s_e v_{qy},
\]
and hence $y_q(t_e) = y_{q,0} + v_{qy} t_e = -N_0 / D(s_e)$. Evaluating~\eqref{eq:fv} at $t_e$ yields $f_{w_j}(t_e) = y_q(t_e)\bigl(d_e - x_r(t_e)\bigr)$, which expands via $1/D(s_e) = y_q(t_e)/(-N_0)$ to
\[
f_{w_j}(t_e) = \frac{y_q(t_e)^2}{-N_0}\Bigl(v_{qy}(s_e d_e) + (v_{qx} - v_{rx})d_e + (v_{rx} y_{q,0} - x_{r,0} v_{qy})s_e + (v_{rx} x_{q,0} - x_{r,0} v_{qx})\Bigr).
\]
Because $y_q(t_e)^2 > 0$ and $-N_0$ has a fixed sign, testing whether $f_{w_j}(t_e) \le 0$ is equivalent to testing whether the fixed point $p(e) = (s_e, d_e, s_e d_e) \in \mathbb{R}^3$ lies in a single halfspace determined solely by the query.

To find the earliest edge $e^*$ of $\gamma^-$ with $t_{e^*} \in (t_{\mathrm{lo}}, t_{\mathrm{hi}})$ and $f_{w_{j^*}}(t_{e^*}) \le 0$, we first isolate the contiguous subchain of $\gamma^-$ whose edges become parallel to $g(t)$ during $(t_{\mathrm{lo}}, t_{\mathrm{hi}})$ via two binary searches on edge slopes using the $\mathrm{up}_j$ pointers, in $\mathcal{O}(\log^2 n)$ time. Along this subchain, the times $t_e$ vary strictly monotonically with the path order. Its interior tree edges in its tree ($T(p^+_{\tau_r})$ or $T(p^-_{\tau_r})$) decompose into $\mathcal{O}(\log n)$ canonical nodes of the biased binary trees~\cite{sleator1983dynamic} together with $\mathcal{O}(\log n)$ off-path edges, plus at most two terminal edges incident to $p^-_{q'}$ and $p^-_{r'}$. We scan these $\mathcal{O}(\log n)$ pieces in increasing order of $t_e$ (forward along $\gamma^-$ if $t_e$ increases along the chain, and in reverse otherwise). Each off-path or terminal edge is tested directly in $\mathcal{O}(1)$ time, while each canonical node is queried via its Dobkin--Kirkpatrick halfspace-emptiness structure~\cite{dobkin1990determining} in $\mathcal{O}(\log n)$ time. This scan takes $\mathcal{O}(\log^2 n)$ time and stops at the first piece containing an edge with $f_{w_j}(t_e) \le 0$. If that piece is a canonical node $\nu^*$, we descend its biased binary tree (of height $\mathcal{O}(\log n)$) in $\mathcal{O}(\log^2 n)$ time: at each step, we query the child corresponding to earlier times in $\mathcal{O}(\log n)$ time, descending into it if it contains a valid edge and into the other child otherwise, until we reach the earliest such edge $e^* = \overline{w_{j^*} w_{j^*+1}}$.

Once $e^*$ is found, let $w^*$ be the endpoint of $e^*$ attaining $G^-(t) = f_{w^*}(t)$ immediately before $t_{e^*}$ (namely, $w_{j^*}$ if $t_e$ increases along $\gamma^-$, and $w_{j^*+1}$ otherwise). By the choice of $e^*$, we have $G^-(t) > 0$ at $t_{\mathrm{lo}}$ and at every earlier breakpoint; since $G^-(t)$ opens downward on each subinterval between consecutive breakpoints, it follows that $G^-(t) > 0$ throughout the entire interval from $t_{\mathrm{lo}}$ up to the start of $w^*$'s interval. Thus $f_{w^*}(t)$ is strictly positive at the start of $w^*$'s interval and satisfies $f_{w^*}(t_{e^*}) \le 0$. Because $f_{w^*}(t)$ is a downward-opening quadratic ($A < 0$), it is positive strictly between its two real roots, so $G^-(t)$ first reaches $0$ at the larger root of $f_{w^*}(t) = 0$, giving $t^*$ in $\mathcal{O}(1)$ time. Finally, if no edge with $t_e \in (t_{\mathrm{lo}}, t_{\mathrm{hi}})$ satisfies $f_{w_j}(t_e) \le 0$, then $G^-(t) > 0$ throughout $[t_{\mathrm{lo}}, t_{\mathrm{hi}}]$ up to the start of the final subinterval ending at $t_{\mathrm{hi}}$, and we test $t_{\mathrm{hi}}$ via an $\mathcal{O}(\log n)$ ray-shooting query. If $\overline{q(t_{\mathrm{hi}})r(t_{\mathrm{hi}})} \subseteq P$, a binary search along $\gamma^-$ matching the left-limit direction of $g(t)$ as $t \to t_{\mathrm{hi}}^-$ locates in $\mathcal{O}(\log^2 n)$ time the vertex $w \in \gamma^-$ attaining $G^-(t) = f_w(t)$ just before $t_{\mathrm{hi}}$, and by the same argument the larger root of $f_w(t) = 0$ yields $t^* \le t_{\mathrm{hi}}$. Thus:

\begin{theorem}
\label{theorem::oneQueryTraj}
Given a fixed trajectory $\tau_r \subset P$, one can build a data structure in $\mathcal{O}(n \log n)$ preprocessing time and $\mathcal{O}(n \log n)$ space so that for any query trajectory $\tau_q \subset P$ and query speeds $v_q$ and $v_r$, the earliest visibility time $t^*$ is computed in $\mathcal{O}(\log^2 n)$ time.
\end{theorem}


\section{Query Trajectories}
\label{sec::queryTrajs}

In this section, we remove all the relaxations in the previous sections, meaning none of $q$'s and $r$'s trajectories are known in advance, along with $v_q$ and $v_r$. Unlike Section~\ref{sec::OneQueryProvidedTrajectory}, where $\tau_r$ is known in advance and can be aligned with the $x$-axis during preprocessing, here neither trajectory is fixed ahead of time. Instead, we parameterize each candidate edge $e$ by its slope and $y$-intercept $(m_e, c_e) \in \mathbb{R}^2$, express the edge-parallel visibility test as a constant-complexity semi-algebraic range query in $\mathbb{R}^2$, and combine the partition tree of Agarwal et al.~\cite{agarwal2013range} with a linear-space compressed range-OR structure.\\

Set $\Delta x_0 := x_{r,0} - x_{q,0}$, $\Delta y_0 := y_{r,0} - y_{q,0}$, $\Delta v_x := v_{rx} - v_{qx}$, and $\Delta v_y := v_{ry} - v_{qy}$, so that $\Delta x(t) := x_r(t) - x_q(t) = \Delta x_0 + \Delta v_x t$ and $\Delta y(t) := y_r(t) - y_q(t) = \Delta y_0 + \Delta v_y t$, and let $M_0 := \Delta v_y \Delta x_0 - \Delta v_x \Delta y_0$. The direction angle $\theta(t)$ of $g(t) = \overrightarrow{q(t)r(t)}$ satisfies $\theta'(t) = M_0 / \|r(t) - q(t)\|^2$. Whenever $\Delta x(t) \neq 0$, the line $g(t)$ has slope $m(t) = \Delta y(t) / \Delta x(t)$ (with $m'(t) = M_0 / \Delta x(t)^2$) and $y$-intercept $c(t) = y_q(t) - m(t) x_q(t)$. Represent each non-vertical edge $e$ with supporting line $y = m_e x + c_e$ by its parameter point $p(e) := (m_e, c_e) \in \mathbb{R}^2$.

\begin{observation}
\label{obs::dualEdgeRange}
On any open interval $(t_1, t_2)$ where $M_0 \neq 0$ and $\Delta x(t) \neq 0$, the set $R_q \subset \mathbb{R}^2$ of parameter points $p(e) = (m_e, c_e)$ of non-vertical edges $e = \overline{w_j w_{j+1}}$ whose parallel instant $t_e$ satisfies $t_e \in (t_1, t_2)$ and $f_{w_j}(t_e) \le 0$ (resp.\ $\ge 0$) is a constant-complexity semi-algebraic range in $\mathbb{R}^2$.
\end{observation}

\begin{proof}
Since $m(t)$ is strictly monotonic on $(t_1, t_2)$, we have $t_e \in (t_1, t_2)$ iff $m_{\min} < m_e < m_{\max}$, where $m_{\min} = \min(m(t_1^+), m(t_2^-))$ and $m_{\max} = \max(m(t_1^+), m(t_2^-))$ (taking the one-sided limits in $\mathbb{R} \cup \{-\infty, +\infty\}$, so the slope range is half-infinite if $\Delta x$ vanishes at $t_1$ or $t_2$). Solving $m(t_e) = m_e$ gives $t_e = (\Delta y_0 - m_e \Delta x_0) / D_x(m_e)$ with $D_x(m_e) := m_e \Delta v_x - \Delta v_y$, and a direct check yields $\Delta x(t_e) D_x(m_e) = -M_0 \neq 0$. Because $w_j$ lies on $y = m_e x + c_e$ and $r(t_e) - q(t_e) = \Delta x(t_e)(1, m_e)$, evaluating~\eqref{eq:fv} gives $f_{w_j}(t_e) = \Delta x(t_e)\bigl(c_e - c(t_e)\bigr)$. Substituting $c(t_e) = (y_{q,0} - m_e x_{q,0}) + (v_{qy} - m_e v_{qx}) t_e$ and $\Delta x(t_e) = -M_0 / D_x(m_e)$ yields
\[
f_{w_j}(t_e) = \frac{-M_0}{D_x(m_e)^2}\Bigl[c_e D_x(m_e) - (y_{q,0} - m_e x_{q,0}) D_x(m_e) - (v_{qy} - m_e v_{qx})(\Delta y_0 - m_e \Delta x_0)\Bigr].
\]
Since $D_x(m_e)^2 > 0$ and $-M_0$ has constant sign on $(t_1, t_2)$, the condition $f_{w_j}(t_e) \le 0$ (resp.\ $\ge 0$) reduces to a single degree-$2$ polynomial inequality in $(m_e, c_e)$ together with $m_{\min} < m_e < m_{\max}$.
\end{proof}

Notice why neither Guibas and Hershberger~\cite{guibas1989optimal} alone nor Section~\ref{sec::OneQueryProvidedTrajectory}'s 3D convex-hull approach achieves $\mathcal{O}(\operatorname{polylog} n)$ query time here in $\mathcal{O}(n)$ space. When the clearing set of a chain forms a single interval (both $\gamma^+$ and $\gamma^-$ when $M_0 = 0$ or $A = 0$, and $\gamma^+$ when $A < 0$ and $M_0 \neq 0$), a local test at an edge $e_v$ in~\cite{guibas1989optimal} suffices to discard one side of the chain. For $\gamma^-$ when $A < 0$ and $M_0 \neq 0$, however, $G^-(t)$ can dip below $0$ on multiple disconnected intervals, so an edge $e_v$ blocked at $t_{e_v}$ gives no directional information about where a clearing edge may lie. Furthermore, Section~\ref{sec::OneQueryProvidedTrajectory}'s 3D reduction relies on aligning the fixed $\tau_r$ with the $x$-axis to eliminate one variable; when both trajectories are queries, the inequality in Observation~\ref{obs::dualEdgeRange} involves all four non-constant monomials $m_e, c_e, m_e^2,$ and $m_e c_e$ (a 4D halfspace query rather than a 3D one, where convex hulls have quadratic worst-case size). Even in $\mathbb{R}^2$, building separate partition trees~\cite{agarwal2013range} across the $\mathcal{O}(\log n)$ levels of $\mathcal{D}$ would use $\mathcal{O}(n \log n)$ space, while reusing shared subtrees as in~\cite{guibas1989optimal} leaves hidden pocket edges inside the subtrees, so a single global partition tree cannot filter only the active edges of a subpolygon unless we also record which ancestors in $\mathcal{D}$ contain each edge.

\medskip
\noindent\textbf{Preprocessing.}
We triangulate $P$ in $\mathcal{O}(n)$ time~\cite{chazelle1991triangulating}, build the $\mathcal{O}(n)$-space ray-shooting structure~\cite{chazelle1994ray}, and build the $\mathcal{O}(n)$-space two-point shortest-path structure of Guibas and Hershberger~\cite{guibas1989optimal} over the balanced subpolygon hierarchy $\mathcal{D}$ of maximum degree $3$ and height $h \in \mathcal{O}(\log n)$~\cite{guibas1989optimal,chazelle1994ray}.

For every pair of bounding diagonals of each $P_u$ ($u \in \mathcal{D}$), we maintain their fundamental hourglass (open or closed) persistently bottom-up via~\cite{guibas1989optimal} in $\mathcal{O}(n)$ total space. An open hourglass has two inward-convex chains (each turning by less than $2\pi$); a closed hourglass has four inward-convex funnel chains (each turning by less than $\pi$) together with a central stem, retained only if it is convex with total turning angle less than $2\pi$ (since no chain of an open query hourglass $H(q', r')$ can traverse a non-convex or spiralling stem). Splitting each retained convex chain of $P_u$ at its at most two vertical transitions yields $\mathcal{O}(1)$ fixed slope-monotone chain types $s$ per node $u$. Starting from the $\mathcal{O}(1)$ boundary edges of each leaf triangle $\Delta$ of $\mathcal{D}$ (with creation node $u_0(e) = \Delta$), each bottom-up concatenation at an internal node $u$ introduces only $\mathcal{O}(1)$ new common-tangent edges (each given a distinct entry with creation node $u_0(e) = u$ so hidden pocket edges never cause false positives). Hence the set $\mathcal{E}$ of all distinct non-vertical edges across $\mathcal{D}$ has size $N = \mathcal{O}(n)$. For each $e \in \mathcal{E}$ and each chain type $s$, we store an $(h + 1)$-bit word $W_s(e)$ ($\mathcal{O}(1)$ machine words) whose $\ell$-th bit is $1$ iff $e$ survives on subchain type $s$ at the unique depth-$\ell$ ancestor of $u_0(e)$ in $\mathcal{D}$. To compute all $W_s(e)$, we visit each $u \in \mathcal{D}$ (at depth $\ell = \operatorname{depth}(u)$), extract the edges of each retained hourglass chain of $P_u$ in $\mathcal{O}(1)$ time per edge using the edge-retrieval lists of~\cite{guibas1989optimal} (Lemma~6), set bit $\ell$ of $W_s(e)$, and discard the temporary list. Since subpolygons at each fixed depth are interior-disjoint, this takes $\mathcal{O}(n \log n)$ time and $\mathcal{O}(n)$ total space.

We preprocess $\{p(e) \mid e \in \mathcal{E}\} \subset \mathbb{R}^2$ into the $d = 2$ partition tree $\mathcal{T}_{\mathcal{E}}$ of Agarwal et al.~\cite{agarwal2013range} in $\mathcal{O}(n^{1+\varepsilon})$ expected time and $\mathcal{O}(n)$ space. By Section~1.1, an in-order leaf traversal of $\mathcal{T}_{\mathcal{E}}$ yields a permutation $A_{\mathcal{E}}[1 \dots N]$ of $\mathcal{E}$ in which every canonical subset or 1D interval returned by a query is a contiguous subarray $A_{\mathcal{E}}[L_j \dots R_j]$, with $M' = \mathcal{O}(n^{1/2}\log^{B_0} n)$ returned intervals in expected time $\mathcal{O}(n^{1/2}\log^{B_0} n)$. Ordering edges created at the same node of $\mathcal{D}$ consecutively in a pre-order traversal of $\mathcal{D}$ assigns each $e \in \mathcal{E}$ a pre-order index $\mathrm{id}(e) \in \{1, \dots, N\}$ so that the descendants of any $u \in \mathcal{D}$ form a contiguous index interval $[\mathrm{id}_L(u), \mathrm{id}_R(u)]$; for any $e$ with $\mathrm{id}(e) \in [\mathrm{id}_L(u), \mathrm{id}_R(u)]$, node $u$ is the unique depth-$\ell$ ancestor of $u_0(e)$ ($\ell = \operatorname{depth}(u)$), so bit $\ell$ of $W_s(e)$ tests survival at $u$. On the points $\{(k, \mathrm{id}(A_{\mathcal{E}}[k])) : 1 \le k \le N\}$, we build a compressed binary range tree splitting on $\mathrm{id}(\cdot)$~\cite{chazelle1988functional}: each node $\nu$ (of size $m_\nu$) stores its $m_\nu$-bit split vector supporting $\mathcal{O}(1)$-time prefix-count queries and a 1D bitwise-OR segment tree of $W_s$ over blocks of size $b = \lceil \log_2 n \rceil$ for each type $s$ (for nodes with $m_\nu < b$, no block tree is built and all $m_\nu$ elements are traced directly), while the values $W_s(e)$ themselves are stored at the leaves. Across all $\lceil \log_2 N \rceil$ levels, the split vectors, block trees, and leaves occupy $\mathcal{O}(n)$ words with $\mathcal{O}(n \log n)$ preprocessing time. Given any query rectangle $[L_j, R_j] \times [\mathrm{id}_L(u), \mathrm{id}_R(u)]$, decomposing $[\mathrm{id}_L(u), \mathrm{id}_R(u)]$ into $\mathcal{O}(\log n)$ tree nodes, querying full blocks via the segment trees in $\mathcal{O}(\log n)$ time per node, and tracing the at most $2b = \mathcal{O}(\log n)$ boundary elements to their leaves in $\mathcal{O}(\log n)$ time each returns the bitwise OR of $W_s(e)$ over the rectangle in $\mathcal{O}(\log^3 n)$ time.

\medskip
\noindent\textbf{Query algorithm.}
Given $\tau_q, \tau_r, v_q,$ and $v_r$, we split the active time interval at the at most two crossings of $\tau_q$ with the supporting line of $\tau_r$ and of $\tau_r$ with the supporting line of $\tau_q$ (handling any collinear subcase as in Section~\ref{sec::OneQueryProvidedTrajectory}). On each resulting subinterval, the active subsegments $s_q \subseteq \tau_q$ and $s_r \subseteq \tau_r$ have endpoints in convex position (or share a single endpoint). Using~\cite{guibas1989optimal} as in Lemma~8 of~\cite{swat2020}, we decide in $\mathcal{O}(\log n)$ time whether $L(s_q, s_r)$ is empty or a single segment, or obtain $q' \subseteq s_q$ and $r' \subseteq s_r$ with $L(s_q, s_r) = H(q', r')$. Restricting to $[t'_0, t'_1]$ where $q(t) \in q'$ and $r(t) \in r'$ (terminating if empty), we check $t'_0$ by $\mathcal{O}(\log n)$ ray shooting (returning $t^* = t'_0$ if visible). Along the triangle path in $\mathcal{D}$ between $p^+_{q'}$ and $p^+_{r'}$ (and symmetrically for $\gamma^-$), concatenating the precomputed fundamental hourglasses of the $\mathcal{O}(\log n)$ traversed subpolygons in $\mathcal{O}(\log^2 n)$ time~\cite{guibas1989optimal} decomposes $\gamma^+$ and $\gamma^-$ in path order into $\mathcal{O}(\log n)$ active slope-monotone subpolygon subchains $(u_i, s_i)$ (each with its active $\partial P$-index interval $[a_i, b_i]$ and slope range $[m_{i,\mathrm{lo}}, m_{i,\mathrm{hi}}]$), together with $\mathcal{O}(\log n)$ dynamic common tangents and terminal segments.

Any standard binary search along a subpolygon chain tree of~\cite{guibas1989optimal} descends deterministically in $\mathcal{O}(\log n)$ time by maintaining the active $\partial P$-index interval $[a_{\mathrm{cur}}, b_{\mathrm{cur}}]$ (the \emph{active-interval descent rule}): at node $v$ with common-tangent edge $e_v$, if $e_v$ is hidden outside $[a_{\mathrm{cur}}, b_{\mathrm{cur}}]$ (tested in $\mathcal{O}(1)$ time by Section~1.1), branch right if $e_v$ lies left of $a_{\mathrm{cur}}$ and left if right of $b_{\mathrm{cur}}$; resolve any zero-length $e_v$ in $\mathcal{O}(1)$ time via the adjacent edge in $v$'s left or right edge-retrieval list; otherwise test $e_v$ and branch accordingly. When $M_0 = 0$ or $A = 0$, inspecting the $\mathcal{O}(\log n)$ dynamic tangents and subpolygon boundary vertices in $\mathcal{O}(\log^2 n)$ time locates the subchains containing the extreme vertices or zeros, and descending each located $\mathcal{O}(\log n)$-height chain tree yields $t^*$ in $\mathcal{O}(\log^2 n)$ time. Similarly, when $A < 0$ and $M_0 \neq 0$ (the case $A > 0$ is symmetric, swapping $\gamma^+$ and $\gamma^-$), the same $\mathcal{O}(\log^2 n)$-time inspection and chain-tree descent along $\gamma^+$ locates $t_{\mathrm{peak}}, t_{\mathrm{lo}},$ and $t_{\mathrm{hi}}$, yielding the single clearing interval $[t_{\mathrm{lo}}, t_{\mathrm{hi}}] \subseteq [t'_0, t'_1]$ of $\gamma^+$.

To search $\gamma^-$ when $A < 0$ and $M_0 \neq 0$ (with $A > 0$ symmetric), we split $[t_{\mathrm{lo}}, t_{\mathrm{hi}}]$ at the at most one vertical-line instant $t_{\mathrm{vert}}$ where $\Delta x(t_{\mathrm{vert}}) = 0$, giving at most two closed subintervals $[t_1, t_2]$ examined in increasing time order. On each $[t_1, t_2]$, we test $t_1$ by $\mathcal{O}(\log n)$ ray shooting (returning $t^* = t_1$ if visible); otherwise, two $\mathcal{O}(\log^2 n)$-time slope queries in~\cite{guibas1989optimal} restrict $\gamma^-$ to the contiguous portion whose directed edge angles lie between $\theta(t_1)$ and $\theta(t_2)$. Since $\theta(t)$ rotates strictly monotonically by less than $\pi$ on $[t_1, t_2]$, every edge on this portion has a unique parallel instant $t_e \in (t_1, t_2)$ varying strictly monotonically along $\gamma^-$. We scan its pieces, namely its $\mathcal{O}(\log n)$ dynamic common tangents, terminal segments, and subpolygon subchains $(u_i, s_i, [m_{i,\mathrm{lo}}, m_{i,\mathrm{hi}}])$, in increasing order of $t_e$ (scanning the path forward if $t_e$ increases along $\gamma^-$, and in reverse if $t_e$ decreases). Dynamic tangents and terminal segments are tested in $\mathcal{O}(1)$ time each. For each subpolygon subchain $(u_i, s_i)$ (at depth $\ell_i = \operatorname{depth}(u_i)$), its active edges are precisely those $e \in \mathcal{E}$ with $\mathrm{id}(e) \in [\mathrm{id}_L(u_i), \mathrm{id}_R(u_i)]$, bit $\ell_i$ of $W_{s_i}(e)$ equal to $1$, and $m_e \in [m_{i,\mathrm{lo}}, m_{i,\mathrm{hi}}]$. Querying $\mathcal{T}_{\mathcal{E}}$ with $R_q \cap \{m_e \in [m_{i,\mathrm{lo}}, m_{i,\mathrm{hi}}]\}$ (Observation~\ref{obs::dualEdgeRange}) returns $M' = \mathcal{O}(n^{1/2}\log^{B_0} n)$ canonical intervals $[L_j, R_j]$, and querying the range-OR structure on $[L_j, R_j] \times [\mathrm{id}_L(u_i), \mathrm{id}_R(u_i)]$ for $j = 1, \dots, M'$ tests bit $\ell_i$ in $\mathcal{O}(n^{1/2}\log^{B_0+3} n)$ expected time.

Scanning at most $\mathcal{O}(\log n)$ subpolygon subchains takes $\mathcal{O}(n^{1/2}\log^{B_0+4} n)$ expected time and stops at the earliest clearing piece. If that piece is a subpolygon subchain $i^*$, we descend the $\mathcal{O}(\log n)$-height chain tree of $u_{i^*}$ using the active-interval descent rule at each unhidden node $v$ (with active slope interval $[m_a, m_b]$ and tangent edge $e_v$):
\begin{itemize}
    \item If $e_v$ has non-zero length, we query $\mathcal{T}_{\mathcal{E}}$ and the range-OR structure on the strictly earlier-$t_e$ slope half excluding $m_{e_v}$ ($[m_a, m_{e_v})$ if $t_e$ increases with slope, or $(m_{e_v}, m_b]$ otherwise); we descend into the earlier-$t_e$ child if bit $\ell_{i^*}$ is $1$, and otherwise test $e_v$ directly (stopping with $e^* = e_v$ if it clears) before descending into the later-$t_e$ child.
    \item If $e_v$ has zero length, we resolve the boundary edge $e_L$ of the earlier-$t_e$ child from $v$'s retrieval list and query the closed earlier-$t_e$ half up to $m_{e_L}$, branching to the earlier-$t_e$ child if bit $\ell_{i^*}$ is $1$ and to the later-$t_e$ child otherwise.
\end{itemize}
This isolates the earliest clearing edge $e^* = \overline{w_{j^*} w_{j^*+1}}$ in $\mathcal{O}(n^{1/2}\log^{B_0+4} n)$ expected time. Solving $f_{w^*}(t) = 0$ for the active endpoint $w^*$ of $e^*$ on the preceding interval inside $[t_1, t_2]$ (as in Section~\ref{sec::OneQueryProvidedTrajectory}) yields $t^*$. Finally, if no interior $t_e \in (t_1, t_2)$ clears, we test $t_2$ via $\mathcal{O}(\log n)$ ray shooting: if $\overline{q(t_2)r(t_2)} \subseteq P$, binary-searching $\gamma^-$ by the left-limit direction of $g(t)$ as $t \to t_2^-$ in $\mathcal{O}(\log^2 n)$ time locates the active vertex $w$ at $t_2$, and the larger root of $f_w(t) = 0$ (as in Section~\ref{sec::OneQueryProvidedTrajectory}) gives $t^* \le t_2$. Setting $B = B_0 + 4$ establishes Theorem~\ref{theorem::queryTrajs}:

\begin{theorem}
\label{theorem::queryTrajs}
Given a simple polygon $P$ with $n$ vertices and any fixed $\varepsilon > 0$, one can build a data structure of size $\mathcal{O}(n)$ in $\mathcal{O}(n^{1+\varepsilon})$ expected preprocessing time so that for any query trajectories $\tau_q, \tau_r \subset P$ and query speeds $v_q$ and $v_r$, the earliest visibility time $t^*$ is computed in $\mathcal{O}(n^{1/2}\log^{B} n)$ expected time, where $B$ is a constant depending on $\varepsilon$.
\end{theorem}


\section{Concluding Remarks and Open Problems}
\label{sec:conclusion}

We presented data structures for computing the first-visibility time $t^*$ for two entities moving along line-segment trajectories inside a simple polygon. Several natural open problems remain. First, extending our results to polygonal domains with $h > 0$ holes: it is unclear whether linear-space sublinear-time queries remain achievable. Second, allowing one or both trajectories to be polylines with $m$ segments, where determining tight trade-offs between $m$ and $n$ for preprocessing and query time is open. Other directions include multi-entity visibility and allowing trajectories to cross $\partial P$.


\section*{AI Disclosure}
The research problem, core ideas, and main results were developed by the authors. AI tools were used for Category~3 (assisting with technical details when revising proofs to address prior reviewer comments), Category~4 (writing and exposition), and Category~5 (checking for errors). All text and proofs were heavily edited and verified by the authors.


\small
\bibliographystyle{abbrv}
\bibliography{bbb}

\end{document}